\documentclass[manuscript]{aastex631}

\usepackage{graphicx}	
\usepackage{amsmath}
\usepackage{amssymb}	
\usepackage{longtable}
\usepackage{threeparttable}
\usepackage{multirow}
\usepackage{booktabs}
\usepackage{array}
\usepackage[figuresright]{rotating}
\usepackage[T1]{fontenc}

\shorttitle{Energy correlations of in- and out-axis GRBs}
\shortauthors{Zhang et al.}

\begin{document}

\title{Spectral Energy Correlations of Gamma-Ray Bursts from Structured Jets}

\author{X. L. Zhang}
\affiliation{School of Physics and Engineering, Qufu Normal University, Qufu 273165, P. R. China\\}
\author{Z. B. Zhang}\thanks{ zbzhang@qfnu.edu.cn}
\affiliation{School of Physics and Engineering, Qufu Normal University, Qufu 273165, P. R. China\\}
\author{Y. F. Huang}\thanks{hyf@nju.edu.cn}
\affiliation{School of Astronomy and Space Science, Nanjing University, Nanjing 210023, P. R. China\\}
\affiliation{Key Laboratory of Modern Astronomy and Astrophysics (Nanjing University), Ministry of Education, P. R. China\\}
\author{D. Li}\thanks{dili@nao.cas.cn}
\affiliation{New Cornerstone Science Laboratory, Department of Astronomy, Tsinghua University, Beijing 100084, China\\}
\affiliation{National Astronomical Observatories, Chinese Academy of Sciences, Beijing 100101, P. R. China\\}
\author{X. J. Li}
\affiliation{School of Cyber Science and Engineering, Qufu Normal University, Qufu 273165, P. R. China}
\author{L. M. Song}
\affiliation{Key Laboratory of Particle Astrophysics, Institute of High Energy Physics, Chinese Academy of Sciences, Beijing 100049, P. R. China}



\begin{abstract}

Using 148 out-axis gamma-ray bursts, we build their spectrum-energy relations of peak energy versus isotropic energy, peak energy versus peak luminosity and peak energy versus jet-calibrated energy which are corrected for a structured jet model. These relations are found to depend on the observer's viewing angle as long as the observer is within the jet cone. After converting the  out-axis energy relations to the in-axis situations, we find that the corresponding in-axis energy relations are universally steeper, of which all of them can be roughly interpreted by the Synchrotron radiation mechanism as shown in Xu et al.. Meanwhile, we notice that the in-axis means of isotropic energies are about one order of magnitude larger than the out-axis means for both short and long bursts except the Supernova-associated gamma-ray bursts. Furthermore, we apply all the newly-found energy relations to construct the Hubble diagrams of out/in-axis bursts. It is found that the in-axis Hubble diagrams are better cosmological indicators.
\end{abstract}



\section{Introduction} \label{sec:intro}

Based on the samples of Gamma-Ray Bursts (GRBs) with known redshifts, people have found many spectrum-energy
	correlations, some of which have been applied for cosmological
	studies. For example, Amati et al. (2002) analyzed 12 BeppoSAX
	(Beppo Satellite per Astronomia X) long GRBs (LGRBs) with the observed peak
	energy ($E_{\rm p}$) and found an empirical relationship between the
	isotropic energy ($E_{\rm{iso}}$) and the intrinsic peak energy
	$E_{\rm pi}=E_{\rm p}(1+z)$ in the rest frame \citep{2002A&A...390...81A}. This energy
	correlation was confirmed by further observations of
	HETE-2 \citep{2004ApJ...602..875S} and BATSE \citep{2002ApJ...576..101L}. \cite{2004ApJ...616..331G} found a tight
	relationship between $E_{\rm pi}$ and the jet-corrected energy
	$E_{\gamma}=(1-cos\theta_{\rm j})E_{\rm iso}$ for relativistic jets with a
	half-opening angle of $\theta_{\rm j}$. Yonetoku et al. (2004) collected
	the data of $E_{\rm pi}$ and peak luminosity ($L_{\rm p}$) observed by
	BeppoSAX and BATSE, and found new relationship between these two
	parameters \citep{2003MNRAS.345..743W,2004ApJ...609..935Y}. With the
	update of observation instruments and the continuous expansion of
	samples with known redshift\citep{2007ApJ...664.1000B,2020ApJ...893...77W}, people
	realized that at least some of the above spectral energy
	relationships also apply to short GRBs (SGRBs) \citep{2012IJMPS..12...19A,2012ApJ...755...55Z}. \cite{2013MNRAS.431.1398T} analyzed the parameters of
	$E_{\rm pi}$, $L_{\rm p}$ and $E_{\rm iso}$ of 13 SGRBs with measured redshifts
	and found $L_{\rm p}\propto E_{\rm p}^{1.59}$ and $E_{\rm iso}\propto
	E_{\rm p}^{1.58}$ for them. \cite{2018PASP..130e4202Z}
	analyzed the above three spectrum-energy correlations for the
	largest GRB sample which includs 31 SGRBs and 252 LGRBs. It is
	found that SGRBs hold similar spectrum-energy relations and have
	almost the same power-law indices as LGRBs. It is very meaningful
	to find out how these spectrum-energy correlations would be if
	GRBs are detected out-axis from us. For this purpose, here we
	collect all out-axis GRB candidates and try to analyze possible
	spectrum-energy correlations.
	
	The above three spectrum-energy relations have been extensively studied
	and applied in cosmology studies \citep{2003MNRAS.345..743W,2020MNRAS.496.1530S}.
	It is well known that Type Ia Supernovae (SNe) can be used as a standard
	candle, but applicable only for redshifts lower than 1.7 currently.
	Undoubtedly, GRBs as one of the farthest objects in cosmos can potentially
	extend the redshift up to $z\sim9.4$ for GRB 090429B \citep{2011ApJ...736....7C},
	and thus can be a complementary tool to SNe Ia. Schaefer et al.(2007)
	constructed a GRB Hubble diagram with 69 GRBs based on the empirical
	$E_{\rm pi}-E_{\rm{\gamma}}$ and $E_{\rm pi}-L_{\rm{p}}$ correlations but found that the error bars are about
	two times more than that of SNe Ia \citep{2007ApJ...660...16S}. In addition, the usage
	of GRBs as cosmological indicators also suffers from some uncertainties
	involving the circularity problem \citep{2019MNRAS.486L..46A} and several observational
	biases \citep{2021MNRAS.501.3515M}, but these problems could be hopefully overcome at
	present with the much expanded sample size. In short, the GRB Hubble diagram
	has been deeply studied by many authors to probe the nature of dark
	energy \citep{2011MNRAS.415.3580D,2021MNRAS.501.3515M} and cosmological
	models \citep{2013MNRAS.431.3301V} in very early stage of the Universe.
	It is worth noting that in previous studies, no one distinguishes between
	in-axis and out-axis GRBs when they are used as cosmological tools. In
	this paper, the GRB Hubble diagrams will be re-constructed for in-axis and
	out-axis GRBs separately, paying special attention on the effect of
	viewing angle ($\theta_{\rm v}$).
	
	GRBs are believed to be produced by relativistic jets. For
	simplicity, the jet was once assumed to be homegeneous and thus
	have a top-hat profile \citep{1999ApJ...525..737R}. In this case, the jet is
	conical and has a uniform distribution of Lorentz factor within a
	half-opening angle $\theta_{\rm j}$. Although the observer may not be
	strictly in-axis \citep{2002ApJ...570L..61G}, the observed properties
	(either the light curve or the spectrum) of GRBs should be very
	similar to those of in-axis case as long as $\theta_{\rm v}<\theta_{\rm j}$ \citep{2000MNRAS.316..943H,2002ApJ...571L..31Y,2018PASP..130e4202Z,2021MNRAS.501.5723F}. However,
	note that the realistic GRB jets should have some complicated
	structures \citep{2003MNRAS.343L..36R}.
	For the structured jets, two different functions are usually assumed
	to describe the profile of
	outflows \citep{2020MNRAS.497.1217T}. One is the
	so called Gaussian jet
	model \citep{2020NatAs...4...77J,2018NatAs...2..751L}, of
	which both the energy density ($\epsilon$) and Lorentz factor ($\Gamma$) are
	normally distributed with respect to the viewing angle. Another
	one is called the power-law jet
	model \citep{2019Sci...363..968G,2001ApJ...552...72D},
	in which the profile of the energy density and Lorentz factor are
	depicted by power-law functions. It is argued that the power-law
	jet model is more favored by various observations \citep{2005ApJ...634.1155X}, so
	we will adopt this picture in our study. Note that for the structured
	jets, the observed GRB properties will be sensitively dependent on
	the viewing angle for an out-axis observer even in the cases of
	$\theta_{\rm v}<\theta_{\rm j}$. In this study, we will investigate this
	effect in depth.
	
	With the development of multi-band observations, a growing number of researchers show great interests in the out-axis properties of GRBs. Particularly, GRB 170817A as the first gravitational wave-associated GRB has been confirmed to be viewed at an angle larger than the half-opening jet angle. It was detected by Fermi \citep{2017ApJ...848L..14G} and found to
	be associated with a LIGO gravitational-wave event GW170817
	\citep{2017ApJ...850L..24G}. The viewing angle $\theta_{\rm v}$ is estimated to be about
	$30$ degrees \citep{2018PTEP.2018d3E02I},
	which is obviously larger than its half-opening angle of
	$\theta_{\rm j}\simeq0.1$ radians \citep{2019ApJ...886L..17H}. \cite{2018PASP..130e4202Z}
	updated the \textbf{$E_{\rm pi}-E_{\rm{iso}}$}, \textbf{$E_{\rm pi}-L_{\rm{p}}$} and \textbf{$E_{\rm pi}-E_{\rm{\gamma}}$} relations with a large
	sample including both short and long GRBs and found that GRB
	170817A is an outlier of these refined relations.
	Recently,  \cite{2020RAA....20..201Z} (hereafter paper I) collected 20 long
	and 22 short GRBs with extended emissions (EEs). They
	concluded that the three spectrum-energy correlations still exist
	for this special kind of bursts. However, GRB 170817A as an EE burst \citep{2021ApJS..252...16L,2022A&A...657A.124L} always
	deviates from these empirical relations no matter whether the out-axis effect is included or not. It indicates that these out-axis GRBs might have a different origin.
	\begin{figure*}
		\centering
		\includegraphics[width=5.5in,height=2.5in]{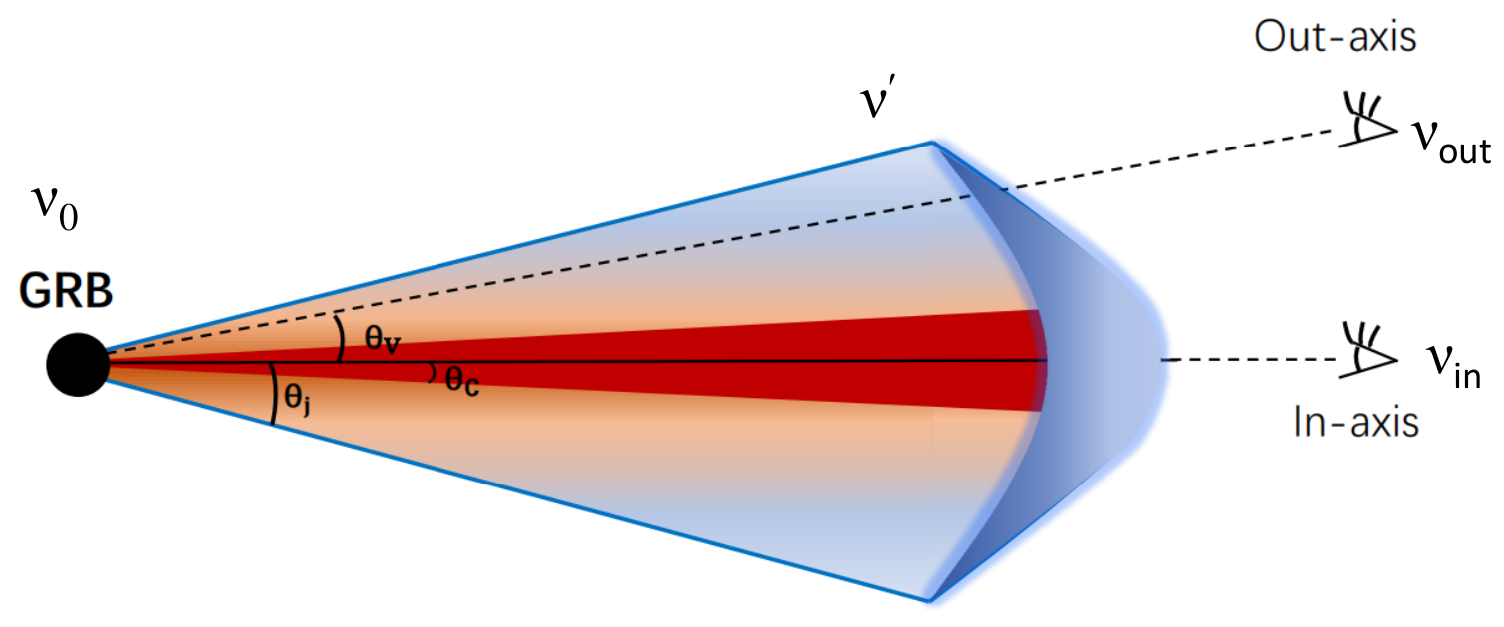}
			\caption{A schematic illustration of the GRB outflow with a
				power-law angular profile viewed in-axis or out-axis by the
				observer. $\theta_{\rm j}$ and $\theta_{\rm c}$ respectively denote the
				half-opening angles of the jet and its central core. $\theta_{\rm v}$ is
				the viewing angle limited to be less than $\theta_{\rm j}$ in the paper. $\nu_0$, $\nu^{'}$ and $\nu_{\rm out}$/$\nu_{\rm in}$ stand for radiation frequencies of photons in the comoving, rest and out/in-axis observer frames, individually. } 
			\label{fig:jet}
		\end{figure*}
		
		Interestingly, the viewing angles of a large sample of GRBs are
		recently derived through Monte-Carlo simulations
		\citep{2015ApJ...799....3R,2019A&A...632A.100H,2021ApJ...907...60M}. We note that the line of sight
		is seldom right on the axis for them. In other words, these events
		could all be regarded as out-axis GRBs, although the line of sight
		is still within the jet range as long as a structured jet is assumed. Note that this is different from the previous out-axis definition for a uniform jet. It is urgent to reveal the
		statistical properties of these out-axis GRBs. Here, we will compare their in-axis parameters with those out-axis ones by assuming a power-law jet (see Figure
		\ref{fig:jet}) and then examine whether some representive spectrum-energy correlations are existent or not in the in/out-axis cases. In addition, the Hubble diagrams of these in/out-axis GRBs will be investigated in details. We take the three cosmological parameters as $H_{0}$=70\
		\textrm{km}\ \textrm{s}$^{-1}$ \textrm{Mpc}$^{-1}$,
		$\Omega_m=0.27$ and $\Omega_\Lambda=0.73$ for the following calculations.
		Our paper is organized as follows. Data and methods are given in Section 2. The main results are presented in Section 3. A summary is made in Section 4.
		
\section{Data and Methods} \label{sec:observations}

\subsection{Sample selection}

	\begin{figure}
	\centering
	\footnotesize
	\includegraphics[height=10cm]{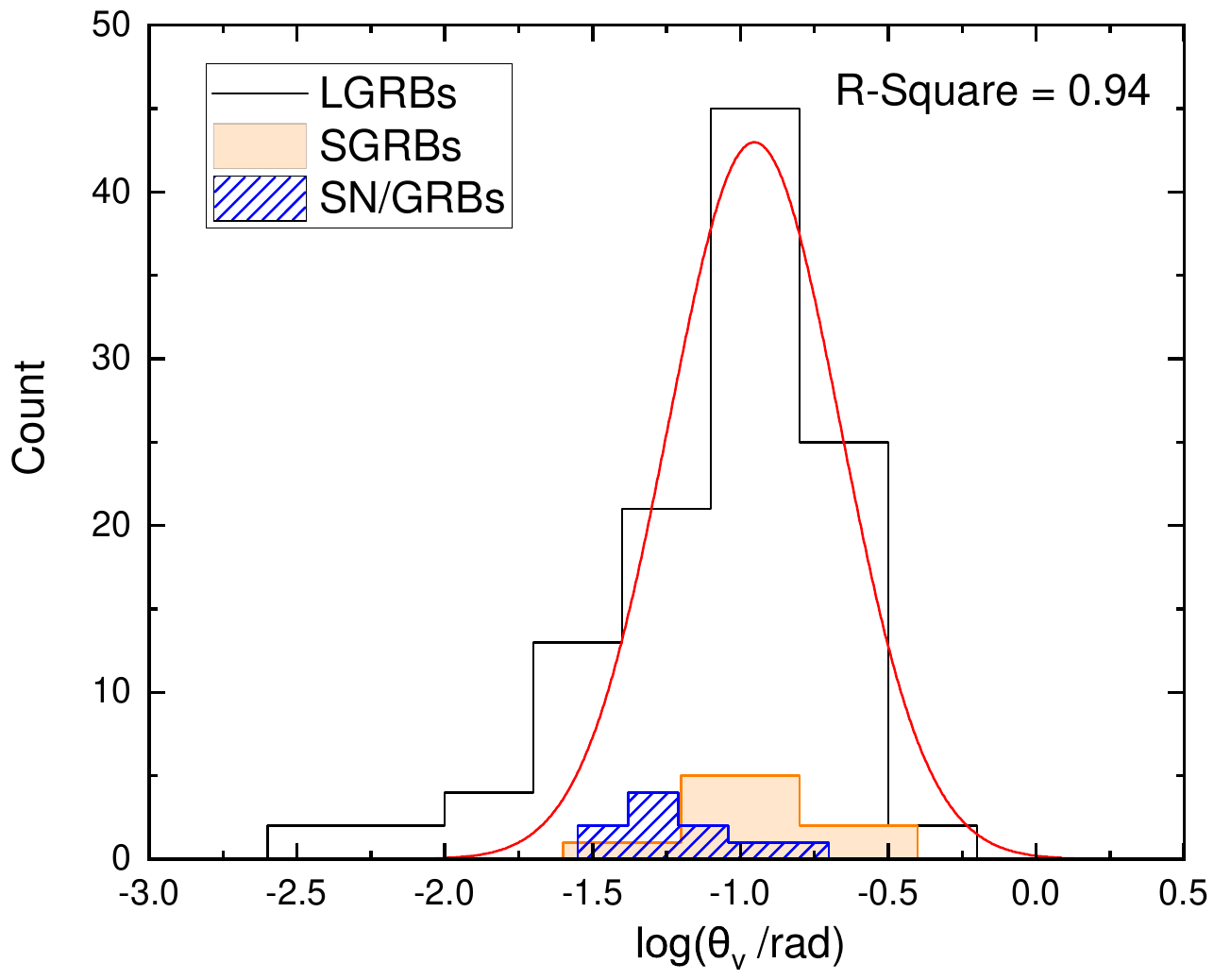}
	\caption{Distributions of the viewing angles ($\theta_{\rm v}$) of 114
		LGRBs (empty), 8 SGRBs (filled) and 10 SN/GRBs (hatched) that have
		been used to calculate the in-axis parameters. The solid red line
		represents the best fit with a lognormal
		function to all the 114 LGRBs with a R-Square value of 0.94.  }
	\label{figs1:thetav}
\end{figure}

\begin{figure}
	\centering
	\footnotesize
	\includegraphics[height=10cm]{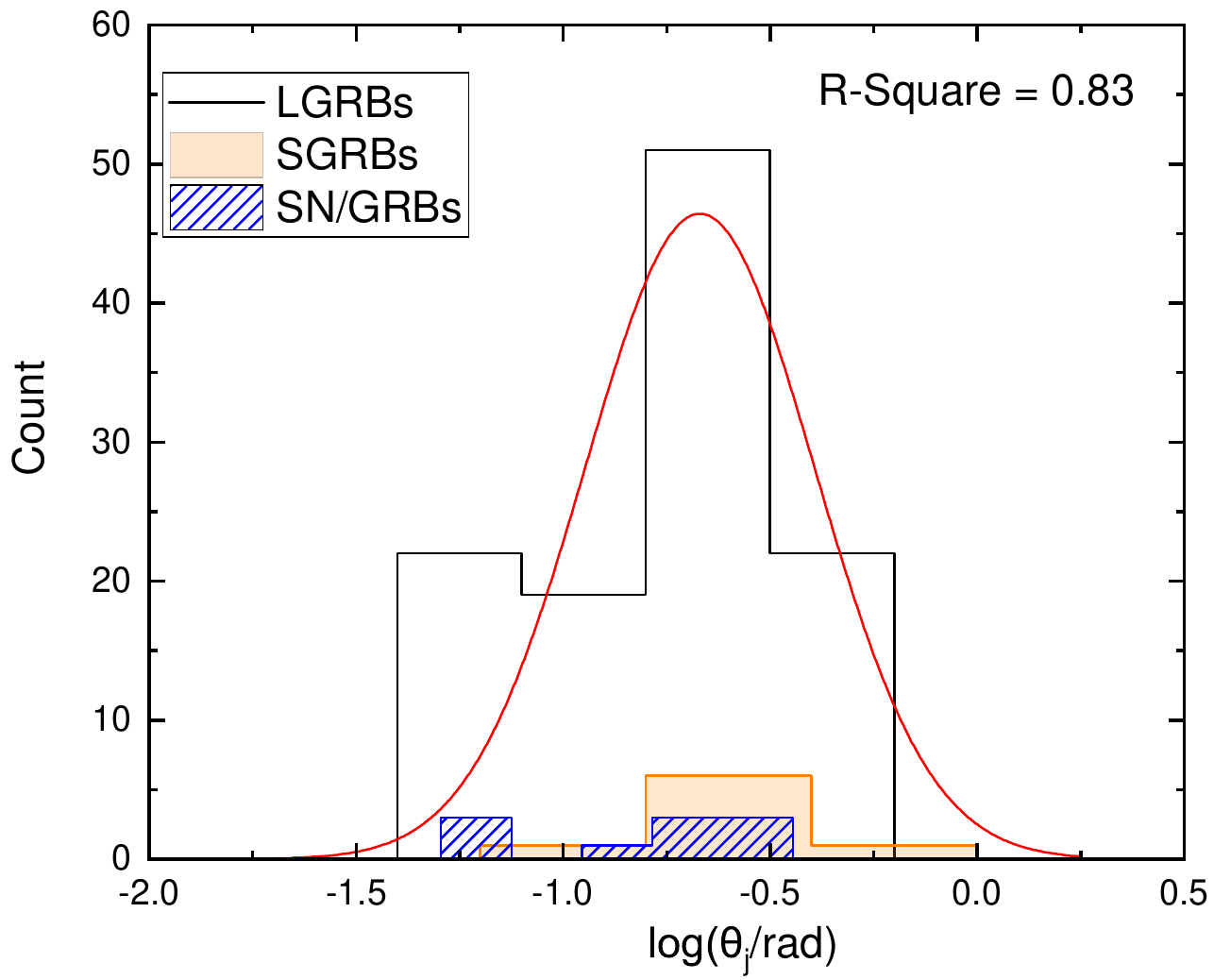}
	\caption{Histograms of the half-opening angles ($\theta_{\rm j}$) of
		different kinds of GRBs. All symbols have the same meanings as in Figure \ref{figs1:thetav}.}
	\label{figs2:thetaj}
\end{figure}
First, we collect several key parameters such as
$E_{\rm p}$, $z$, $\theta_{\rm j}$ and $\theta_{\rm v}$ in the literature \citep[e.g.][]{2015ApJ...799....3R,2019A&A...632A.100H,2019MNRAS.489.2104T}, and constructed a sample of 148 out-axis GRB candidates  including 128 LGRBs, 8 SGRBs and 12 SN/GRBs. These events were recorded between January 2005 and August 2017. Among them, there are 111 GRBs detected by Swift/BAT and 37 GRBs detected by Fermi/GBM. It is worth
noting that all the jet and viewing angles of these GRBs in our sample were taken from literature \citep{2015ApJ...799....3R,2019A&A...632A.100H,2020MNRAS.497.4672A}, where they used a ScaleFit software package to perform the Markov-Chain-Monte-Carlo (MCMC) simulation on the multi-wavelength afterglows under an assumption of a top-hot jet. It proves a structured jet is also available for the $\theta_{\rm v}$ simulations \citep[e.g.][]{2019MNRAS.489.2104T}. In practice, a realistic jet is most probably structured \citep{2020ApJ...894...11G}. If the resulting multi-wavelength spectra and light curves in theory do not rely on the geometry of the jets seriously within the narrow jet ejecta, the inferred jet angles and viewing angles should be comparable between different types of jet structures. The physical parameters of our sample are listed in Table \ref{tab:sample}. Figure \ref{figs1:thetav} displays the lognormal distribution of $\theta_{\rm v}$,	with a mean value of $- 1.06$ (0.4 dex), $ -0.85$ (0.23 dex) and $ -1.20$ (0.22 dex) for LGRBs, SGRBs and SN/GRBs, respectively. We fit the distribution of viewing angles with a lognormal function and get the best fit with a R-square of 0.94, indicating a perfect fit of the model to the data. Figure \ref{figs2:thetaj} shows the lognormal distribution of $\theta_{\rm j}$,  with a mean value of $-0.75$ (0.30 dex), $-0.54$ (0.16 dex) and $-0.92$ (0.27 dex) for LGRBs, SGRBs and SN/GRBs, correspondingly. Similarly, we apply a lognormal function to fit the distribution of half-opening angles and get the best statistic of R-square=0.83. The initial Lorentz factor ($\Gamma$) is obtained on basis of the break time of X-ray afterglow \citep{2006ApJ...642..389N}. In total, $\Gamma$ is available for 114 long, 8 short and 10 SN-associated	GRBs (see the following Sec 2.3). Finally, the observed peak energies of these three types of GRBs	are on average $150.9\pm43.2$ keV, $158.9\pm31.8$ keV and $124.0\pm20.7$ keV, respectively.

	\setlength{\tabcolsep}{0.7mm}
	\setlength{\LTcapwidth}{4in}
	\small
	\begin{longtable}{lcccccccccccc}
		\caption{Physical parameters of out-axis GRBs }
		\label{tab:sample}\\
		
		\toprule
		GRB& T$_{\rm 90}$ & $z$ & $E_{\rm p}$ & $\alpha$  & $S_{\rm \gamma}$ & $P_{\rm \gamma}$ & $\Delta E$ & $\theta_{\rm j}$ & $\theta_{\rm v}$ &$\Gamma$& $Ref$ \\
		& $(\rm s)$ & & $(\rm keV)$ & & $(\rm erg\ cm^{-2})$ &$(\rm ph\ cm^{-2}\ s^{-1})$ & $(\rm keV)$ &(rad)& (rad)&&\\
		(1)&(2)&(3)&(4)&(5)&(6)&(7)&(8)&(9)&(10)&(11)&(12)\\
		\midrule
		\endfirsthead
		
		\endhead

		\bottomrule
		\endlastfoot
		
		050126 & 24.8 & 1.29 & 93.75$\pm$26.13 & -1.35 & $(8.40\pm1.09)\times10^{-7}$ & 1.30$\pm$0.10 & 15-150 & 0.38 & 0.16 & -- & [1,7,4]\\
		050315$^{a}$$^{b}$ & 95.4 & 1.95 & 33.87$\pm$13.15 & -1.77 & $(3.08\pm0.17)\times10^{-6}$ & 1.85$\pm$0.20 & 15-150 & 0.34 & 0.06 & 59.63 & [1,7,4]\\
		050318$^{a}$ & 32.0 & 1.44 & 49.23$\pm$9.44  & -1.03 & $(1.04\pm0.08)\times10^{-6}$ & 3.00$\pm$0.20 & 15-150 & 0.15 & 0.05 & 66.53 & [1,7,4]\\
		050319$^{a}$ & 151.6 & 3.24 & 44.73$\pm$1.62  & -1.62 & $(1.29\pm0.17)\times10^{-6}$ & 1.38$\pm$0.20 & 15-150 & 0.05 & 0.03 & 65.16 & [1,7,4]\\
		050401$^{a}$ & 32.1 & 2.90 & 120.00$\pm$20.00 & -1.39 & $(8.04\pm0.37)\times10^{-6}$ & 12.10$\pm$1.20 & 15-150 & 0.46 & 0.33 & 131.75 & [1,7,4]\\
		050416A$^{a}$ & 2.50 & 0.65 & 14.85$\pm$6.97  & -0.82 & $(3.74\pm0.50)\times10^{-7}$ & 4.40$\pm$0.80 & 15-150 & 0.24 & 0.08 & 111.65 & [1,7,4] \\
		050505$^{a}$ & 58.9 & 4.27 & 131.95$\pm$45.00 & -1.40 & $(2.45\pm0.19)\times10^{-6}$ & 1.79$\pm$0.30 & 15-150 & 0.19 & 0.08 & 109.62 & [1,7,4] \\
		050525A$^{a}$ & 8.84 & 0.61 & 80.40$\pm$3.48  & -1.78 & $(1.51\pm0.02)\times10^{-5}$ & 27.10$\pm$3.00 & 15-150 & 0.06 & 0.03 & 111.43 & [1,7,4] \\
		050802$^{a}$ & 19.0 & 1.71 & 175.60$\pm$85.56 & -1.48 & $(2.19\pm0.20)\times10^{-6}$ & 0.85$\pm$0.08 & 15-150 & 0.29 & 0.16 & 105.80 & [1,7,4] \\
		050820A$^{a}$$^{b}$ & 240.8 & 2.61 & 100.00$\pm$15.00 & -1.24 & $(3.86\pm0.23)\times10^{-6}$ & 4.70$\pm$0.30 & 15-150 & 0.15 & 0.09 & 44.42 & [1,7,4] \\
		050824$^{a}$ & 22.60 & 0.83 & 80.00$\pm$8.00 & -1.00 & $(2.56\pm0.45)\times10^{-7}$ & 0.22$\pm$0.03 & 15-150 & 0.14 & 0.05 & 44.98 & [1,7,4] \\
		050904 & 174.2 & 6.29 & 314.00$\pm$173.00 & -1.23 & $(1.40\pm0.26)\times10^{-5}$ & 0.26$\pm$0.01 & 15-5000 & 0.15 & 0.07 & -- & [1,7,4] \\
		050908 & 19.4 & 3.35 & 50.46$\pm$9.61 & -1.83 & $(4.42\pm0.59)\times10^{-7}$ & 0.31$\pm$0.03 & 15-150 & 0.25 & 0.08 & -- & [1,7,4] \\
		050922C$^{a}$$^{b}$ & 4.55 & 2.20 & 209.57$\pm$76.97 & -1.36 & $(1.60\pm0.01)\times10^{-6}$ & 2.92$\pm$0.10 & 15-150 & 0.07 & 0.05 & 210.88 & [1,4] \\
		051016B$^{a}$ & 4.00 & 0.94 & 20.42$\pm$2.00 & -1.59 & $(1.04\pm0.01)\times10^{-7}$ & 1.26$\pm$0.16 & 15-150 & 0.34 & 0.20 & 84.11 & [1,7,4] \\
		051109A$^{a}$ & 37.2 & 2.35 & 50.50$\pm$11.38 & -0.76 & $(2.09\pm0.29)\times10^{-6}$ & 3.82$\pm$0.67 & 15-150 & 0.29 & 0.17 & 97.39 & [1,7,4] \\
		051111$^{a}$ & 46.1 & 1.55 & 179.70$\pm$54.52 & -1.32 & $(4.22\pm0.17)\times10^{-6}$ & 0.54$\pm$0.02 & 15-150 & 0.28 & 0.14 & 73.01 & [1,4] \\
		060115$^{a}$ & 139.6 & 3.53 & 69.81$\pm$15.07 & -1.01 & $(1.71\pm0.15)\times10^{-6}$ & 0.90$\pm$0.10 & 15-150 & 0.18 & 0.14 & 70.16 & [1,7,4] \\
		060124$^{a}$ & 13.4 & 2.30 & 29.64$\pm$3.66 & -1.40 & $(1.43\pm0.24)\times10^{-5}$ & 0.84$\pm$0.18 & 20-2000 & 0.21 & 0.12 & 150.10 & [1,7,4] \\
		060206$^{a}$ & 7.60 & 4.05 & 80.74$\pm$38.83 & -1.12 & $(8.42\pm0.44)\times10^{-7}$ & 2.78$\pm$0.20 & 15-150 & 0.38 & 0.12 & 193.67 & [1,7,4] \\
		060210$^{a}$$^{b}$ & 255.0 & 3.91 & 150.10$\pm$58.00 & -1.19 & $(7.56\pm0.44)\times10^{-6}$ & 2.70$\pm$0.30 & 15-150 & 0.06 & 0.04 &77.34& [1,7,4]\\
		060306$^{a}$ & 60.9 & 3.50 & 67.17$\pm$11.66 & -1.76 & $(2.05 \pm 0.14) \times 10^{-6}$ & 0.45$\pm$0.03 & 15-150 & 0.08 & 0.07 & 104.94 & [1,4] \\
		060418$^{a}$ & 103.1 & 1.49 & 134.99$\pm$101.64 & -1.55 & $(8.31 \pm 0.29) \times 10^{-6}$ & 4.75$\pm$0.50 & 15-150 & 0.31 & 0.14 & 70.89 & [1,7,4] \\
		060502A$^{a}$ & 28.4 & 1.51 & 151.50$\pm$50.00 & -1.12 & $(2.29 \pm 0.11) \times 10^{-6}$ & 1.70$\pm$0.20 & 15-150 & 0.32 & 0.14 & 84.51 & [1,7,4] \\
		060512 & 8.50 & 0.44 & 19.04$\pm$2.00 & -1.58 & $(2.03 \pm 0.31) \times 10^{-7}$ & 0.90$\pm$0.20 & 15-150 & 0.27 & 0.15 & -- & [1,7,4] \\
		060522$^{a}$$^{b}$ & 71.1 & 5.11 & 73.31$\pm$16.00 & -0.70 & $(1.04 \pm 0.13) \times 10^{-6}$ & 0.50$\pm$0.15 & 15-150 & 0.33 & 0.12 & 113.38 & [1,7,4] \\
		060526$^{a}$ & 298.2 & 3.21 & 227.80$\pm$61.00 & -1.97 & $(1.22 \pm 0.19) \times 10^{-6}$ & 1.60$\pm$0.18 & 15-150 & 0.05 & 0.01 & 55.33 & [1,7,4] \\
		060604$^{a}$ & 95.0 & 2.14 & 74.51$\pm$7.00 & -2.12 & $(3.79 \pm 0.98) \times 10^{-7}$ & 0.60$\pm$0.10 & 15-150 & 0.06 & 0.05 & 67.61 & [1,7,4] \\
		060605$^{a}$ & 79.1 & 3.80 & 87.04$\pm$19.90 & -0.63 & $(7.67 \pm 1.56) \times 10^{-7}$ & 0.50$\pm$0.10 & 15-150 & 0.05 & 0.01 & 78.71 & [1,7,4] \\
		060607A$^{a}$$^{b}$ & 102.2 & 3.08 & 137.53$\pm$39.99 & -1.15 & $(2.51 \pm 0.13) \times 10^{-6}$ & 1.40$\pm$0.10 & 15-150 & 0.37 & 0.22 & 77.53 & [1,7,4] \\
		060614$^{a}$ & 108.7 & 0.13 & 134.10$\pm$24.20 & -1.82 & $(1.88 \pm 0.15) \times 10^{-5}$ & 11.40$\pm$0.70 & 15-150 & 0.29 & 0.11 & 26.42 & [1,7,4] \\
		060707$^{a}$ & 66.2 & 3.43 & 60.94$\pm$9.27 & -0.42 & $(1.58 \pm 0.15) \times 10^{-6}$ & 1.10$\pm$0.20 & 15-150 & 0.22 & 0.13 & 83.95 & [1,7,4] \\
		060714$^{a}$$^{b}$ & 115.0 & 2.71 & 56.85$\pm$13.00 & -1.54 & $(2.77 \pm 0.21) \times 10^{-6}$ & 1.20$\pm$0.10 & 15-150 & 0.06 & 0.04 & 76.99 & [1,7,4] \\
		060814$^{a}$ & 145.3 & 0.84 & 257.00$\pm$58.00 & -1.43 & $(2.69 \pm 0.12) \times 10^{-5}$ & 7.23$\pm$0.30 & 20-2000 & 0.36 & 0.13 & 49.16 & [1,7,4] \\
		060904B$^{a}$$^{b}$ & 189.9 & 0.70 & 82.48$\pm$22.41 & -1.66 & $(1.64 \pm 0.18) \times 10^{-6}$ & 0.11$\pm$0.01 & 15-150 & 0.08 & 0.06 & 33.52 & [1,4] \\
		060906$^{a}$ & 44.6 & 3.69 & 47.80$\pm$18.97 & -1.98 & $(2.20 \pm 0.16) \times 10^{-6}$ & 0.72$\pm$0.04 & 15-150 & 0.05 & 0.03 & 118.15 & [1,4] \\
		060908$^{a}$ & 19.3 & 1.88 & 127.72$\pm$24.75 & -1.32 & $(2.81 \pm 0.11) \times 10^{-6}$ & 1.20$\pm$0.05 & 15-150 & 0.38 & 0.15 & 126.73 & [1,4] \\
		060926 & 8.00 & 3.20 & 4.21$\pm$0.40 & -1.92 & $(2.08 \pm 0.15) \times 10^{-7}$ & 1.10$\pm$0.10 & 15-150 & 0.28 & 0.14 & -- & [1,7,4] \\
		060927$^{a}$ & 22.5 & 5.60 & 70.90$\pm$9.94 & -0.81 & $(1.12 \pm 0.07) \times 10^{-6}$ & 2.68$\pm$0.20 & 15-150 & 0.31 & 0.16 & 158.26 & [1,7,4] \\
		061007 & 59.0 & 1.26 & 561.00$\pm$27.00 & -0.70 & $(4.50 \pm 0.05) \times 10^{-5}$ & 14.30$\pm$0.30 & 100-1000 & 0.32 & 0.17 & -- & [1,4] \\
		061222A$^{a}$$^{b}$ & 71.4 & 2.09 & 233.69$\pm$69.96 & -1.02 & $(2.66 \pm 0.26) \times 10^{-5}$ & 7.50$\pm$0.23 & 20-2000 & 0.07 & 0.04 & 92.91 & [1,7,4] \\
		070419A & 115.6 & 0.97 & 26.48$\pm$7.14 & -0.84 & $(5.82 \pm 1.09) \times 10^{-7}$ & 1.40$\pm$0.20 & 15-150 & 0.18 & 0.12 & -- & [1,7,4] \\
		070506$^{a}$ & 4.30 & 2.31 & 31.03$\pm$3.58 & -0.77 & $(2.07 \pm 0.32) \times 10^{-7}$ & 0.90$\pm$0.10 & 15-150 & 0.25 & 0.13 & 146.48 & [1,7,4] \\
		070508$^{a}$ & 20.9 & 0.82 & 233.00$\pm$12.00 & -0.96 & $(2.00 \pm 0.03) \times 10^{-5}$ & 24.20$\pm$0.60 & 100-1000 & 0.45 & 0.34 & 99.57 & [1,7,4] \\
		070521$^{a}$ & 37.9 & 0.55 & 195.00$\pm$123.00 & -1.10 & $(8.01 \pm 1.96) \times 10^{-6}$ & 6.50$\pm$0.26 & 15-150 & 0.15 & 0.06 & 58.76 & [1,7,4] \\
		070611$^{a}$ & 12.2 & 2.04 & 54.89$\pm$10.33 & -0.75 & $(3.72 \pm 0.75) \times 10^{-7}$ & 0.60$\pm$0.10 & 15-150 & 0.27 & 0.11 & 83.83 & [1,7,4] \\
		070714B$^{a}$ & 64.0 & 0.92 & 164.87$\pm$91.74 & -0.86 & $(7.20 \pm 0.90) \times 10^{-7}$ & 2.70$\pm$0.20 & 15-150 & 0.33 & 0.28 & 50.14 & [1,4] \\
		070724A$^{a}$ & 0.43 & 0.46 & 45.85$\pm$17.12 & -0.68 & $(3.01 \pm 0.52) \times 10^{-8}$ & 0.94$\pm$0.17 & 15-150 & 0.28 & 0.14 & 93.18 & [1,4] \\
		070802$^{a}$ & 16.4 & 2.45 & 58.31$\pm$5.80 & -1.04 & $(2.59 \pm 0.57) \times 10^{-7}$ & 0.30$\pm$0.14 & 15-150 & 0.48 & 0.15 & 93.14 & [1,7,4] \\
		070809$^{a}$ & 1.28 & 0.22 & 81.87$\pm$49.45 & -1.43 & $(7.33 \pm 1.13) \times 10^{-8}$ & 0.97$\pm$0.12 & 15-150 & 0.13 & 0.05 & 65.63 & [1,6] \\
		070810A$^{a}$ & 11.0 & 2.17 & 42.23$\pm$6.46 & -1.37 & $(6.29 \pm 0.62) \times 10^{-7}$ & 1.85$\pm$0.20 & 15-150 & 0.10 & 0.08 & 126.30 & [1,7,4] \\
		071003$^{a}$ & 150.0 & 1.10 & 799.00$\pm$100.00 & -0.97 & $(8.30 \pm 0.30) \times 10^{-6}$ & 6.30$\pm$0.40 & 15-150 & 0.30 & 0.18 & 55.04 & [1,4] \\
		071010A$^{a}$ & 6.00 & 0.98 & 36.40$\pm$4.00 & -0.62 & $(3.61 \pm 0.53) \times 10^{-7}$ & 0.90$\pm$0.20 & 15-150 & 0.30 & 0.18 & 75.57 & [1,7,4] \\
		071010B & 36.12 & 0.97 & 57.33$\pm$5.81 & -1.52 & $(4.37 \pm 0.08) \times 10^{-6}$ & 7.40$\pm$0.30 & 15-150 & 0.34 & 0.11 & -- & [1,7,4] \\
		071020$^{a}$ & 4.20 & 2.14 & 322.00$\pm$53.00 & -0.65 & $(2.30 \pm 0.10) \times 10^{-6}$ & 8.40$\pm$0.30 & 15-150 & 0.21 & 0.19 & 192.82 & [1,4] \\
		071122 & 68.7 & 1.14 & 138.90$\pm$69.00 & -1.42 & $(6.07 \pm 1.28) \times 10^{-7}$ & 0.40$\pm$0.20 & 15-150 & 0.25 & 0.13 & -- & [1,7,4] \\
		080310$^{a}$ & 365.0 & 2.43 & 23.28$\pm$11.36 & -1.65 & $(2.26 \pm 0.25) \times 10^{-6}$ & 1.30$\pm$0.20 & 15-150 & 0.10 & 0.03 & 47.71 & [1,7,4] \\
		080319B$^{a}$ & 124.9 & 0.94 & 651.00$\pm$13.00 & -0.82 & $(8.10 \pm 0.10) \times 10^{-5}$ & 24.80$\pm$0.50 & 15-150 & 0.10 & 0.06 & 73.06 & [1,4] \\
		080319C$^{a}$ & 34.0 & 1.95 & 158.79$\pm$52.43 & -0.67 & $(6.40 \pm 0.40) \times 10^{-6}$ & 1.35$\pm$0.40 & 100-1000 & 0.30 & 0.14 & 119.01 & [1,7,4] \\
		080411$^{a}$ & 56.0 & 1.03 & 259.00$\pm$27.00 & -1.51 & $(2.64 \pm 0.01) \times 10^{-5}$ & 43.20$\pm$0.90 & 15-150 & 0.10 & 0.07 & 79.82 & [1,4] \\
		080413A$^{a}$& 46.0 & 2.43 & 170.00$\pm$80.00& -1.20 & $(3.49 \pm 0.12) \times 10^{-6}$ & 5.40$\pm$0.20 & 15-150 & 0.26 & 0.14 &111.07& [1,7,4]\\
		080413B$^{a}$& 8.00 & 1.10 &  69.00$\pm$7.91 & -1.23 & $(3.14 \pm 0.08) \times 10^{-6}$ & 18.70$\pm$0.80& 15-150 & 0.14 & 0.05 &121.89& [1,7,4]\\
		080430$^{a}$$^{b}$& 16.2 & 0.77 &  6.13$\pm$1.00  & -1.73 & $(1.19 \pm 0.03) \times 10^{-6}$ & 2.65$\pm$0.17 & 15-150 & 0.06 & 0.03 &73.97& [1,7,4]\\
		080605$^{a}$$^{b}$& 20.0 & 1.64 & 333.00$\pm$34.00& -0.94 & $(1.34 \pm 0.02) \times 10^{-5}$ & 19.70$\pm$0.60& 20-2000& 0.36 & 0.18 &124.71& [1,7,4]\\
		080721$^{a}$$^{b}$& 16.2 & 2.60 & 485.00$\pm$59.00& -0.93 & $(1.20 \pm 0.10) \times 10^{-5}$ & 20.90$\pm$1.80& 15-150 & 0.11 & 0.08 &199.99& [1,4]\\
		080905B$^{a}$& 105.9& 2.37 & 181.20$\pm$60.71& -0.86 & $(2.91 \pm 0.04) \times 10^{-6}$ & 4.08$\pm$1.10 & 10-1000& 0.16 & 0.09 &69.28& [2,7,4]\\
		080913 & 8.00 & 6.44 & 98.20$\pm$22.89 & -0.39 & $(5.54 \pm 0.60) \times 10^{-7}$ & 0.70$\pm$0.10 & 15-150 & 0.36 & 0.14 &--& [1,7,4]\\
		080916A$^{a}$& 46.3 & 0.69 & 107.57$\pm$18.52& -0.82 & $(7.81 \pm 0.08) \times 10^{-6}$ & 7.10$\pm$1.35 & 10-1000& 0.16 & 0.12 &47.99 & [2,7,4]\\
		080928$^{a}$ &280.0 & 1.69 & 52.24$\pm$5.00  & -1.73 & $(2.46 \pm 0.13) \times 10^{-6}$ & 2.10$\pm$0.15 & 15-150 & 0.25 & 0.16 &45.45 & [1,7,4]\\
		081007$^{a}$ & 9.73 & 0.53 & 27.25$\pm$11.52 & -1.52 & $(5.79 \pm 0.44) \times 10^{-7}$ & 2.90$\pm$0.40 & 15-150 & 0.16 & 0.12 &66.78 & [1,7,4]\\		
		081008$^{a}$ &185.5 & 1.97 & 105.25$\pm$29.00& -1.35 & $(4.15 \pm 0.20) \times 10^{-6}$ & 1.26$\pm$0.10 & 15-150 & 0.06 & 0.02 &56.19 & [1,7,4]\\
		081028$^{a}$ &260.0 & 3.04 & 70.74$\pm$13.88 & -1.40 & $(3.80 \pm 0.23) \times 10^{-6}$ & 0.56$\pm$0.13 & 15-150 & 0.30 & 0.16 &58.01 & [1,7,4]\\
		081029$^{a}$ &270.0 & 3.85 & 149.70$\pm$75.00& -1.10 & $(2.10 \pm 0.27) \times 10^{-6}$ & 0.40$\pm$0.17 & 15-150 & 0.16 & 0.01 &61.23 & [1,7,4]\\
		081121 & 41.9 & 2.51 & 160.80$\pm$16.56& -0.43 & $(1.53 \pm 0.02) \times 10^{-5}$ & 12.81$\pm$1.66& 10-1000& 0.28 & 0.21 &--& [2,7,4]\\
		081203A$^{a}$&294.0 & 2.10 &201.38$\pm$126.49& -1.33 & $(7.70 \pm 0.30) \times 10^{-6}$ & 2.90$\pm$0.20 & 15-150 & 0.15 & 0.06 &55.47 & [1,4]\\
		081221$^{a}$ & 29.7 & 2.26 & 86.94$\pm$1.41  & -0.90 & $(3.00 \pm 0.09) \times 10^{-5}$ & 27.48$\pm$1.36& 10-1000& 0.34 & 0.11 &150.60& [2,7,4]\\
		081222$^{a}$ & 18.9 & 2.70 & 142.70$\pm$9.58 & -0.86 & $(1.19 \pm 0.01) \times 10^{-5}$ & 14.50$\pm$1.00& 10-1000& 0.08 & 0.02 &129.39& [2,7,4]\\
		090102$^{a}$ & 28.3 & 1.55 & 451.00$\pm$58.00& -1.24 & $(7.04 \pm 0.35) \times 10^{-6}$ & 5.50$\pm$0.80 & 20-2000& 0.38 & 0.29 &80.88 & [1,7,4]\\
		090205$^{a}$$^{b}$& 8.80 & 4.70 & 38.42$\pm$9.92  & -0.39 & $(1.74 \pm 0.32) \times 10^{-7}$ & 0.40$\pm$0.10 & 15-150 & 0.05 & 0.01 &149.86& [1,7,4]\\
		090323$^{a}$ &135.17& 3.57 & 632.90$\pm$40.83& -1.29 & $(1.18 \pm 0.01) \times 10^{-4}$ & 14.33$\pm$0.84& 10-1000& 0.31 & 0.16 &100.21& [2,7,4]\\
		090328A& 61.7 & 0.74 & 639.70$\pm$45.71& -1.09 & $(4.40 \pm 0.01) \times 10^{-5}$ & 25.35$\pm$1.50& 10-1000& 0.31 & 0.17 &--& [2,7,4]\\
		090418A$^{a}$& 56.0 & 1.61 &610.00$\pm$164.00& -1.30 & $(4.60 \pm 0.20) \times 10^{-6}$ & 1.90$\pm$0.30 & 15-150 & 0.07 & 0.02 &91.00 & [1,4]\\
		090423$^{a}$ & 7.17 & 8.20 & 50.55$\pm$5.66  & -0.59 & $(8.16 \pm 0.72) \times 10^{-7}$ & 4.24$\pm$1.22 & 10-1000& 0.26 & 0.13 &254.65& [2,7,4]\\
		090424$^{a}$$^{b}$& 14.1 & 0.54 & 154.00$\pm$3.83 & -1.04 & $(4.63 \pm 0.01) \times 10^{-5}$ &126.70$\pm$2.04& 10-1000& 0.22 & 0.16 &71.62 & [2,7,4]\\
		090426$^{a}$ & 1.24 & 2.61 & 55.09$\pm$16.19 & -1.11 & $(1.76 \pm 0.29) \times 10^{-7}$ & 2.01$\pm$0.26 & 15-150 & 0.31 & 0.15 &259.53& [1,7,4]\\
		090510$^{a}$ & 0.30 &   0.90 &366.93$\pm$149.24& -0.80 & $(3.40 \pm 0.40) \times 10^{-7}$ & 9.70$\pm$1.10 & 15-150 & 0.39 & 0.22 &263.65& [1,4]\\
		090516$^{a}$$^{b}$&123.1 & 4.10 & 142.10$\pm$26.45& -1.52 & $(1.72 \pm 0.06) \times 10^{-5}$ & 7.54$\pm$1.10 & 10-1000& 0.07 & 0.02 &103.16& [2,7,4]\\
		090529$^{a}$ & 70.4 & 2.63 & 42.07$\pm$8.45  & -0.87 & $(9.34 \pm 1.25) \times 10^{-7}$ & 0.70$\pm$0.20 & 15-150 & 0.48 & 0.07 &60.97 & [1,7,4]\\
		090618$^{a}$ & 113.3& 0.54 & 141.25$\pm$14.52& -1.49 & $(1.09 \pm 0.01) \times 10^{-4}$ & 6.72$\pm$0.13 & 15-150 & 0.06 & 0.04 &59.27 & [1,4]\\
		090715B$^{a}$&266.0 & 3.00 & 134.00$\pm$30.00& -1.10 & $(5.70 \pm 0.20) \times 10^{-6}$ & 3.80$\pm$0.20 & 15-150 & 0.31 & 0.16 &67.18 & [1,4]\\
		090726$^{a}$ & 67.0 & 2.71 & 26.88$\pm$2.00  & -1.30 & $(7.87 \pm 0.90) \times 10^{-7}$ & 0.70$\pm$0.15 & 15-150 & 0.17 & 0.07 &66.72 & [1,7,4]\\
		090809$^{a}$ & 5.40 & 2.74 & 198.00$\pm$13.00& -0.85 & $(3.40 \pm 0.50) \times 10^{-7}$ & 1.10$\pm$0.20 & 8-1000 & 0.31 & 0.15 &130.54& [1,4]\\
		090812$^{a}$ &66.7  & 2.45 &271.83$\pm$154.38& -1.03 & $(5.80 \pm 0.20) \times 10^{-6}$ & 3.60$\pm$0.20 & 15-150 & 0.29 & 0.20 &100.24& [1,4]\\
		090926A$^{a}$& 13.76& 2.11 & 339.80$\pm$5.75 & -0.86 & $(1.47 \pm 0.01) \times 10^{-4}$ &135.50$\pm$2.01& 10-1000& 0.32 & 0.20 &168.35& [3,4]\\
		090926B& 55.6 & 1.24 & 82.49$\pm$3.00  & -0.52 & $(1.08 \pm 0.01) \times 10^{-5}$ & 6.31$\pm$1.00 & 10-1000& 0.35 & 0.14 &--& [2,7,4]\\
		090927$^{a}$ & 2.16 &   1.37 & 61.95$\pm$19.12 & -1.30 & $(2.00 \pm 0.30) \times 10^{-7}$ & 2.00$\pm$0.20 & 15-150 & 0.45 & 0.26 &130.27& [1,7,4]\\
		091003$^{a}$ & 20.2 & 0.90 & 367.30$\pm$26.76& -1.07 & $(2.33 \pm 0.01) \times 10^{-5}$ & 46.63$\pm$2.21& 10-1000& 0.31 & 0.18 &71.53 & [2,7,4]\\
		091018$^{a}$ & 4.40 & 0.97 &    19.43$\pm$2.00 & -1.76 & $(1.37 \pm 0.04) \times 10^{-6}$ &  9.80$\pm$0.40& 15-150 & 0.30 & 0.17 &141.10& [1,7,4]\\
		091020$^{a}$ & 24.3 & 1.71 & 244.20$\pm$36.72& -1.25 & $(8.35 \pm 0.15) \times 10^{-6}$ & 10.30$\pm$1.27& 10-1000& 0.28 & 0.18 &94.34 & [2,7,4]\\
		091024$^{a}$ & 93.9 & 1.09 &349.03$\pm$176.80& -1.33 & $(8.56 \pm 0.06) \times 10^{-6}$ & 5.65$\pm$1.17 & 10-1000& 0.31 & 0.16 &56.79 & [2,7,4]\\
		091029$^{a}$ & 39.2 & 2.75 & 59.35$\pm$8.01  & -1.46 & $(2.41 \pm 0.11) \times 10^{-6}$ & 1.70$\pm$0.20 & 15-150 & 0.13 & 0.10 &100.70& [1,7,4]\\
		091109A$^{a}$& 48.0 & 3.50 & 63.93$\pm$6.00  & -1.32 & $(1.65 \pm 0.21) \times 10^{-6}$ & 1.20$\pm$0.40 & 15-150 & 0.22 & 0.10 &115.91& [1,7,4]\\
		091127$^{a}$ & 8.70 & 0.49 & 35.49$\pm$1.54  & -1.26 & $(2.07 \pm 0.01) \times 10^{-5}$ &103.00$\pm$2.22& 10-1000& 0.15 & 0.14 &92.20 & [2,7,4]\\
		091208B$^{a}$$^{b}$&12.5 & 1.06 & 38.45$\pm$5.75  & -0.15 & $(6.19 \pm 0.19) \times 10^{-6}$ & 31.00$\pm$1.43& 10-1000& 0.10 & 0.02 &112.97& [2,7,4]\\
		100219A$^{a}$& 18.8 & 4.70 & 129.10$\pm$65.00& -0.89 & $(4.50 \pm 0.82) \times 10^{-7}$ & 0.40$\pm$0.10 & 15-150 & 0.05 & 0.003&121.94& [1,7,4]\\
		100302A$^{a}$& 17.9 & 4.81 & 90.17$\pm$39.00 & -1.43 & $(3.07 \pm 0.45) \times 10^{-7}$ & 0.50$\pm$0.10 & 15-150 & 0.19 & 0.07 &131.51& [1,7,4]\\
		100316B$^{a}$& 3.80 & 1.18 & 16.31$\pm$2.00  & -1.80 & $(2.04 \pm 0.15) \times 10^{-7}$ & 1.30$\pm$0.20 & 15-150 & 0.15 & 0.12 &123.65& [1,7,4]\\
		100418A$^{a}$& 7.93 & 0.62 & 187.32$\pm$85.54& -2.16 & $(3.40 \pm 0.50) \times 10^{-7}$ & 1.00$\pm$0.20 & 15-150 & 0.24 & 0.05 &75.34 & [1,4]\\
		100424A$^{a}$& 104.0& 2.47 & 95.25$\pm$73.00 & -1.69 & $(1.48 \pm 0.14) \times 10^{-6}$ & 0.40$\pm$0.10 & 15-150 & 0.06 & 0.03 &83.52 & [1,7,4]\\
		100425A$^{a}$& 37.0 & 1.76 & 25.35$\pm$9.28  & -0.89 & $(4.61 \pm 0.94) \times 10^{-7}$ & 1.40$\pm$0.20 & 15-150 & 0.16 & 0.07 &66.52 & [1,7,4]\\
		100615A$^{a}$& 37.4 & 1.40 & 53.35$\pm$9.93  & -0.90 & $(8.72 \pm 0.08) \times 10^{-6}$ & 10.12$\pm$0.96& 10-1000& 0.28 & 0.13 &79.31 & [2,7,4]\\
		100621A$^{a}$& 63.60& 0.54 & 68.19$\pm$6.00  & -1.81 & $(2.04 \pm 0.04) \times 10^{-5}$ & 12.60$\pm$0.40& 15-150 & 0.05 & 0.03 &53.45 & [1,7,4]\\
		100724A$^{a}$& 1.39 & 1.29 & 42.50$\pm$8.20  & -0.51 & $(1.41 \pm 0.22) \times 10^{-7}$ & 1.55$\pm$0.19 & 15-150 & 0.28 & 0.14 &139.62& [1,7,4]\\
		100728A$^{a}$$^{b}$&165.4& 2.11 & 290.00$\pm$7.82 & -0.64 & $(1.28 \pm 0.01) \times 10^{-4}$ & 13.03$\pm$1.20& 10-1000& 0.11 & 0.08 &87.63& [2,7,4]\\
		100728B$^{a}$& 10.24& 1.57 & 109.80$\pm$22.04& -0.78 & $(3.34 \pm 0.06) \times 10^{-6}$ & 8.42$\pm$1.27 & 10-1000& 0.30 & 0.17 &123.78& [2,7,4]\\
		100816A$^{a}$& 2.05 & 0.80 & 126.70$\pm$7.53 & -0.23 & $(3.65 \pm 0.05) \times 10^{-6}$ & 19.88$\pm$1.08& 10-1000& 0.29 & 0.13 &148.53& [2,7,4]\\
		100906A$^{a}$$^{b}$&110.6& 1.73 & 69.67$\pm$10.10 & -0.90 & $(2.33 \pm 0.01) \times 10^{-5}$ & 19.42$\pm$1.85& 10-1000& 0.05 & 0.02 &70.83& [2,7,4]\\
		101219A$^{a}$& 0.60 & 0.72 & 490.00$\pm$79.00& -0.63 & $(4.60 \pm 0.30) \times 10^{-7}$ & 4.10$\pm$0.20 & 15-150 & 0.29 & 0.14 &177.56& [1,4]\\
		110128A& 12.16& 2.34 &192.80$\pm$112.90& -1.26 & $(1.43 \pm 0.10) \times 10^{-6}$ & 2.84$\pm$1.05 & 10-1000& 0.47 & 0.07 &--& [2,7,4]\\
		110205A$^{a}$& 257.0& 1.98 & 105.10$\pm$19.23& -1.52 & $(3.36 \pm 0.35) \times 10^{-5}$ & 4.00$\pm$0.20 & 20-2000& 0.39 & 0.27 &70.20 & [2,7,4]\\
		110213A$^{a}$& 34.31& 1.46 & 74.73$\pm$13.07 & -1.42 & $(9.37 \pm 0.05) \times 10^{-6}$ & 21.63$\pm$2.32& 10-1000& 0.29 & 0.15 &94.99 & [2,7,4]\\
		110422A$^{a}$$^{b}$&25.9 & 1.77 & 109.39$\pm$6.29 & -0.83 & $(3.83 \pm 0.05) \times 10^{-5}$ & 28.60$\pm$1.00& 15-150 & 0.07 & 0.05 &136.78& [1,7,4]\\
		110503A$^{a}$& 10.0 & 1.61 & 129.32$\pm$25.71& -0.88 & $(1.12 \pm 0.05) \times 10^{-5}$ & 29.60$\pm$1.30& 15-150 & 0.40 & 0.30 &135.61& [1,7,4]\\
		110715A$^{a}$& 13.0 & 0.82 &  89.27$\pm$6.02 & -1.25 & $(1.12 \pm 0.01) \times 10^{-5}$ & 52.40$\pm$1.20& 15-150 & 0.34 & 0.13 &120.81& [1,7,4]\\
		110731A$^{a}$& 7.49 & 2.83 & 319.30$\pm$19.69& -0.87 & $(2.29 \pm 0.01) \times 10^{-5}$ & 29.11$\pm$2.11& 10-1000& 0.18 & 0.13 &232.77& [2,7,4]\\
		110801A$^{a}$$^{b}$&385.0& 1.86 & 83.45$\pm$29.00 & -1.61 & $(4.45 \pm 0.32) \times 10^{-6}$ & 1.00$\pm$0.20 & 15-150 & 0.42 & 0.21 &43.47 & [1,7,4]\\
		110808A$^{a}$& 48.0 & 1.35 &    31.80$\pm$3.00 & -1.35 & $(3.36 \pm 1.05) \times 10^{-7}$ & 0.60$\pm$0.20 & 15-150 & 0.19 & 0.06 &49.03& [1,7,4]\\
		110818A$^{a}$& 67.1 & 3.36 & 183.40$\pm$85.30& -1.12 & $(5.15 \pm 0.03) \times 10^{-6}$ & 4.88$\pm$1.50 & 10-1000& 0.23 & 0.14 &90.76 & [2,7,4]\\
		111008A$^{a}$& 63.5 & 4.99 & 149.00$\pm$28.00& -1.36 & $(5.15 \pm 0.29) \times 10^{-6}$ & 6.40$\pm$0.70 & 20-2000& 0.05 & 0.03 &126.16& [2,7,4]\\
		111107A$^{a}$& 12.03& 2.89 & 108.00$\pm$32.00& -1.38 & $(9.07 \pm 0.35) \times 10^{-7}$ & 4.84$\pm$1.88 & 15-150 & 0.28 & 0.12 &150.10& [1,7,4]\\
		111228A$^{a}$& 99.8 & 0.72 &  26.51$\pm$1.25 & -1.58 & $(1.81 \pm 0.01) \times 10^{-5}$ & 27.58$\pm$1.74& 10-1000& 0.05 & 0.03 &51.44 & [2,7,4]\\
		120118B$^{a}$& 23.3 & 2.94 & 36.89$\pm$11.84 & -1.59 & $(1.67 \pm 0.13) \times 10^{-6}$ & 2.14$\pm$0.30 & 15-150 & 0.29 & 0.13 &124.41& [1,7,4]\\
		120119A$^{a}$& 55.3 & 1.73 & 183.30$\pm$7.96 & -0.96 & $(3.87 \pm 0.01) \times 10^{-5}$ & 22.37$\pm$1.71& 10-1000& 0.05 & 0.02 &88.48 & [2,7,4]\\
		120327A$^{a}$& 63.5 & 2.81 & 106.09$\pm$22.36& -1.36 & $(3.53 \pm 0.14) \times 10^{-6}$ & 0.52$\pm$0.02 & 15-150 & 0.41 & 0.12 &102.04& [1,4]\\
		120422A$^{a}$& 5.35 & 0.28 & 97.01$\pm$9.00  & -1.19 & $(3.01 \pm 0.99) \times 10^{-7}$ & 0.56$\pm$0.15 & 15-150 & 0.20 & 0.09 &46.25 & [1,7,4]\\
		120711A$^{a}$& 44.0 & 1.41 & 973.00$\pm$35.00& -0.94 & $(1.94 \pm 0.01) \times 10^{-4}$ & 26.70$\pm$0.60& 10-1000& 0.25 & 0.16 &111.52& [2,4]\\
		120712A$^{a}$& 22.6 & 4.17 & 101.38$\pm$19.81& -0.60 & $(4.43 \pm 0.05) \times 10^{-6}$ & 5.51$\pm$1.11 & 10-1000& 0.30 & 0.26 &122.38& [3,7,4]\\
		120802A$^{a}$& 50.0 & 3.80 & 52.96$\pm$6.84  & -1.22 & $(1.64 \pm 0.13) \times 10^{-6}$ & 2.98$\pm$0.23 & 15-150 & 0.21 & 0.12 &103.59& [1,7,4]\\
		120811C$^{a}$& 24.3 & 2.67 & 47.31$\pm$4.71  & -1.33 & $(2.84 \pm 0.11) \times 10^{-6}$ & 1.14$\pm$0.04 & 15-150 & 0.23 & 0.15 &125.36& [1,4]\\
		120815A& 9.70 & 2.36 & 27.25$\pm$3.00  & -1.18 & $(4.63 \pm 0.43) \times 10^{-7}$ & 2.20$\pm$0.30 & 15-150 & 0.32 & 0.13 &--& [2,7,4]\\
		120907A$^{a}$$^{b}$&5.76 & 0.97 & 122.40$\pm$34.15& -0.81 & $(8.09 \pm 0.41) \times 10^{-7}$ & 7.56$\pm$1.94 & 10-1000& 0.12 & 0.03 &104.67& [2,7,4]\\
		120909A$^{a}$&115.0 & 3.93 & 335.00$\pm$25.00& -1.23 & $(7.04 \pm 0.42) \times 10^{-6}$ & 3.10$\pm$0.20 & 20-2000& 0.22 & 0.15 &89.48 & [2,7,4]\\
		121027A$^{a}$& 62.6 & 1.77 & 61.75$\pm$13.25 & -1.58 & $(1.95 \pm 0.18) \times 10^{-6}$ & 1.00$\pm$0.20 & 15-150 & 0.06 & 0.04 &58.41 & [1,7,4]\\
		121128A$^{a}$& 17.3 & 2.20 & 64.92$\pm$5.39  & -1.00 & $(9.30 \pm 0.11) \times 10^{-6}$ & 26.18$\pm$2.35& 10-1000& 0.27 & 0.14 &147.26& [2,7,4]\\
		121211A$^{a}$& 5.63 & 1.02 & 100.20$\pm$15.84& -0.27 & $(6.41 \pm 0.40) \times 10^{-7}$ & 4.30$\pm$1.07 & 10-1000& 0.30 & 0.18 &92.54 & [2,7,4]\\
		121229A$^{a}$& 111.5& 2.71 & 27.55$\pm$3.00  & -1.32 & $(7.75 \pm 2.25) \times 10^{-7}$ & 0.44$\pm$0.23 & 15-150 & 0.28 & 0.15 &60.12 & [1,7,4]\\
		170817A&2.00 &0.0098& 214.70$\pm$56.60& -0.60 & $(2.79 \pm 0.17) \times 10^{-7}$ & 3.70$\pm$0.80 &10-1000 & 0.10 & 0.53 &8& [2,7]\\
		171205A& 189.4 & 0.04 & 120.56$\pm$85.84 & -0.88 & $(3.53 \pm 0.28) \times 10^{-6}$ & 0.95$\pm$0.28 & 15--150 &0.60 &0.73 &109.6& [1,6,8]\\
		
	\end{longtable}
	
	\footnotesize
	Note-(1) Column 1 gives the GRB names, Column 2 provides the duration ($T_{\rm 90}$), Column 3 lists the cosmological redshifts, Column 4 gives the observed $E_{\rm p}$. Column 5 gives the low energy spectral indexes $\alpha$ of the $\nu$F$_\nu$ spectra, with mean values being -1.18, -0.84 and -1.28 for LGRBs, SGRBs and SN/GRBs, respectively. Columns 6 and 7 are the observed energy fluences ($S_{\rm \gamma}$) and peak photon fluxes ($P_{\rm \gamma}$), in units of erg\ $\rm cm^{-2}$ and ph\ $\rm cm^{-2}$\ s$^{-1}$, respectively. Column 8 provides the energy bands ($\Delta E$). Columns 9 and 10 show the viewing angle ($\theta_{\rm v}$) and the half-opening jet angle ($\theta_{\rm j}$). Column 11 displays the Lorentz factors $\Gamma$ of the jet. Finally, reference for the spectral parameters ($E_{\rm p}$, $\alpha$,  $S_{\rm \gamma}$ and $P_{\rm \gamma}$), $\theta_{\rm v}$ and $\theta_{\rm j}$ are listed in Column 12.\\
	(2) References are given in order for $T_{\rm 90}$, $z$, $E_{\rm p}$, $\alpha$, $S_{\rm \gamma}$, $P_{\rm \gamma}$, $\Delta E$, $\theta_{\rm j}$ and $\theta_{\rm v}$ as follows: [1]. \textit{Swift} official website at https://swift.gsfc.nasa.gov/archive/grb\_table/;
	[2]. https://heasarc.gsfc.nasa.gov/W3Browse/fermi/fermigtrig.html;
	[3]. http://gcn.gsfc.nasa.gov/; [4]\cite{2015ApJ...799....3R};
	[5]. \cite{2019A&A...632A.100H}; [6]. \cite{2021ApJ...907...60M};
	[7]. \cite{2018PASP..130e4202Z}; \\
	$^a$ 132 GRBs with both $\theta_v$ and $\Gamma$ available have been applied to
	study the in-axis spectrum-energy correlations. \\
	$^b$ The viewing angles of these GRBs are well constrained by the MCMC simulations.\\
	
\clearpage
	
	\subsection{Conversion of parameters from out-axis to in-axis directions}
	As illustrated in Figure \ref{fig:jet}, we define the frequencies of prompt $\gamma$-rays in the
	comoving (central engine) frame and the rest (ejecta) frame as $\nu_0$ and $\nu^{'}$, respectively. The
	observed frequencies are denoted as $\nu_{\rm out}$ in the out-axis case and $\nu_{\rm in}$
	in the in-axis case. Note that the in-axis concept
	here strictly refers to the situation that the observer is right located on the
	jet axis with an viewing angle of $\theta_{\rm v}=0$ degree \citep{1998ApJ...497L..17S}, which is
	different from the traditional on-axis definition of $\theta_{\rm v}<\theta_{\rm j}$ for
	top-hat jets by many other authors \citep[e.g.][hereafter X23]{2000MNRAS.316..943H,2002ApJ...570L..61G,2021MNRAS.501.5723F,2023A&A...673A..20X}. The observed features are almost identical as long
	as the observer is within the range of the homogeneous jet core with $0 \leq \theta_{\rm v} \leq \theta_{\rm c}$. Considering the effects of cosmological dilation and special relativity, the
	relation between the above frequencies can be expressed as
	\begin{equation}\label{equation:1}
		\nu^{'}=\nu_{\rm out}(1+z)=\frac{\nu_0}{\Gamma(1-\beta{cos}\theta_{\rm v})}=\nu_0\delta,
	\end{equation}
	where $\beta=(1-1/\Gamma^2)^{{1}/{2}}$ denotes the ratio of outflow velocity to the speed of light,
	$\delta$ is the Doppler factor, $\theta_{\rm v}$ is the out-axis viewing angle between the line of sight and the jet axis.
	Then one can get
	\begin{equation}
		\label{equation:2}
		\nu_{0}=\frac{\nu_{\rm out}(1+z)}{\delta}.
	\end{equation}
In case of a jet viewed in-axis, we can obtain $\theta_{\rm v}\simeq0$, $\nu_{out}=\nu_{in}$ and  $\delta\simeq2\Gamma$, leading to $\nu_{0}=\nu_{in}(1+z)\Gamma^{-1}/2$. 
	
	Many authors pointed out that a realistic jet
	is likely to be structured \cite[e.g.][]{2019MNRAS.489.2104T,2020ApJ...894...11G}. Therefore, it is reasonable to utilize the structured jet model to study the impirical energy relations. Here, we assume that the realitivistic jet has a power-law form \citep{2001ApJ...552...72D,2002ApJ...571..876Z,2002MNRAS.332..945R,2003A&A...400..415W}. In other words, the energy density $\epsilon$ (defined as the energy per unit solid angle)
	and the Lorentz factor vary with the polar angle $\theta$ ($\equiv\theta_v$) as
	\begin{equation}
		\label{equation:3}
		\epsilon=\left\{
		\begin{aligned}
			\epsilon_c \ \ \ \ \ \ \ \ \ (0\leq\theta\leq\theta_c), \\
			\epsilon_c(\theta/\theta_c)^{-2}\ \ \  (\theta_c<\theta\leq\theta_j),
		\end{aligned}
		\right.
	\end{equation}
	and
	\begin{equation}
		\label{equation:4}
		\Gamma=\left\{
		\begin{aligned}
			\Gamma_c \ \ \ \ \ \ \ \ \ (0\leq\theta\leq\theta_c), \\
			\Gamma_c(\theta/\theta_c)^{-\kappa}\ \ \  (\theta_c<\theta\leq\theta_j),
		\end{aligned}
		\right.
	\end{equation}
	within the jet angle,
	where $\theta_c$ and $\Gamma_c$ are respectively the angle and the Lorentz factor of jet core.
	Here the jet core is assumed to be a miniature top-hat jet to avoid the divergence on the jet
	axis ($\theta=0$), i.e. the outflow material is uniformly distributed inside the central core.
	
	For simplicity, we follow \cite{2002MNRAS.332..945R} to take $\theta_c=3^\circ$ and
	$\kappa=2$ for all bursts in our sample. Note that $\kappa=2$ is the upper limit of the
	range of 1.5--2 for the power law index, as suggested by \cite{2002ApJ...571..876Z}. Thus the observed isotropic energy
	and Lorentz factor for an observer at a viewing angle of $\theta_v$ can be written as
	$E_{\text{iso,out}}=4\pi D_L^2S_{bol}(1+z)^{-1}$ and $\Gamma_{\text{out}}=\Gamma$.
	We also assume the two in-axis parameters to hold the similar power-law relations as
	\begin{equation}
		\label{equation:5}
		E_{\text{iso,in}}=\left\{
		\begin{aligned}
			E_{\text{iso,out}} \ \ \ \ \ \ \ \ \ (0\leq\theta\leq\theta_c), \\
			E_{\text{iso,out}}(\theta/\theta_c)^{2}\ \ \  (\theta_c<\theta\leq\theta_j),
		\end{aligned}
		\right.
	\end{equation}
	and
	\begin{equation}
		\label{equation:6}
		\Gamma_{\text in}\equiv\Gamma_c=\left\{
		\begin{aligned}
			\Gamma_{\text{out}} \ \ \ \ \ \ \ \ \ (0\leq\theta\leq\theta_c), \\
			\Gamma_{\text{out}}(\theta/\theta_c)^{2}\ \ \  (\theta_c<\theta\leq\theta_j).
		\end{aligned}
		\right.
	\end{equation}
	Similarly, the observed peak luminosity for an out-axis observer is related with the in-axis one by
	\begin{equation}
		\label{equation:7}
		L_{\text{p,in}}=\left\{
		\begin{aligned}
			L_{\text{p,out}} \ \ \ \ \ \ \ \ \ (0\leq\theta\leq\theta_c), \\
			L_{\text{p,out}}(\theta/\theta_c)^{2}\ \ \  (\theta_c<\theta\leq\theta_j),
		\end{aligned}
		\right.
	\end{equation}
	in which \textbf{$L_{\text{p,out}}=4\pi D_L^2(z)P_{\text{bol}}$ }represents the observed peak luminosity at a viewing angle of $\theta_v$. The jet-corrected energy, $E_{\gamma}=f_bE_{\rm iso}$, can be obtained by a beaming factor of $f_b=(1-cos\theta_j)$. Hence, the out/in-axis $E_{\gamma}$ are correlated with
\begin{equation}
	\label{equation:27}
	E_{\gamma,\text{in}}=\left\{
	\begin{aligned}
		E_{\gamma,\text{out}} \ \ \ \ \ \ \ \ \ (0\leq\theta\leq\theta_c), \\
		E_{\gamma,\text{out}}(\theta/\theta_c)^{2}\ \ \  (\theta_c<\theta\leq\theta_j),
	\end{aligned}
	\right.
\end{equation}

	From Eq. (\ref{equation:2}), the observed
	out/in-axis frequencies are $\nu_{\text{out}}\simeq\nu_0\cdot2\Gamma_{\text{out}}/(1+z)$ and $\nu_{\text{in}}\simeq\nu_0\cdot2\Gamma_{\text{in}}/(1+z)$. Hence, the intrinsic peak
	energy in the rest frame is
	\begin{equation}
		\label{equation:8}
		E_{\text {pi,in}}\equiv\frac{\nu_{\text in}}{\nu_{\text{out}}}E_{\text{pi,out}}\simeq\frac{\Gamma_{\text in}}{\Gamma_{\text{out}}}E_{\text{pi,out}}.
	\end{equation}
	
	We can convert the observed out-axis quantities to the corresponding in-axis ones
	via Eqs. (\ref{equation:5}-\ref{equation:8}) when $0\leq\theta_v\leq\theta_j$
	is satisfied. It is noticeable that the in-axis and out-axis observations are
	thought to be undistinguishable when the jet core is viewed
	frontally ($\theta_v\leq\theta_c$). Undoubtedly, for an observer with $\theta_v>\theta_j$,
	the spectrum-energy relations will be altered drastically. Unfortunately, there is only one burst satisfying the condition, that is GRB 170817A whose viewing angle and jet angle are about 42 and 34 degrees, respectively \citep{2024ApJ...962..117L}. GRB 170817A is not involved in the current analysis since the curvature effect \citep{2004ApJ...614..284D,2007AN....328...99Z} and the edge effect \citep{1999Natur.398..389K} of its wider jet will play an improtant role on the observations.
	
	\subsection{Constraining the Lorentz factors with X-ray afterglows}
As shown in Equation (\ref{equation:2}), the initial Lorentz factor of GRBs plays an important role in the relationship	between the out- and in-axis parameters. To derive the Lorentz factor $\Gamma$, a smoothly broken power-law function \citep{2019ApJS..245....1T} is used to fit the X-ray afterglows of GRBs in our sample, i.e.,
\begin{equation}\label{equation:9}
		F_X(t)=F_{X0}\left[\left(\frac{t}{t_{b}}\right)^{\alpha_1\omega}+\left(\frac{t}{t_{b}}\right)^{\alpha_2\omega}\right]^{-1/\omega},
	\end{equation}
where $\alpha_1$ and $\alpha_2$ are power-law indices during the shallow decay phase and the normal decay phase, respectively, $\omega$ is a smoothness parameter assumed to be $\sim$3 here,	$t_{b}$ represents the observed end time of the X-ray plateau phase, and $2^{-1/\omega}F_{X0}$ describes the observed flux at the time $t_b$. 

According to \cite{1976PhFl...19.1130B} and \cite{2006ApJ...642..389N}, the Lorentz factor of outflows should evolve with the observational time of $t$ as
\begin{equation}\label{equation:11}
	\Gamma(t)=\left[\frac{(17-4k)E_{\rm iso}(1+z)^{3-k}}{16C^{3-k}\pi{c}^{5-k}t^{3-k}}\right]^{1/2(4-k)},
\end{equation}
where $k$ is the power-law index of ambient density as a function of the radius ($R$),
i.e., $n=n_0R^k$, $C$ and $k$ are connected with $C=4(4-k)/(5-4)$ at the outer edge of the external
shock \citep{2002ApJ...568..820G}, and $E_{iso}$ is the isotropic energy in prompt $\gamma$-rays. Letting $t_{\rm day}=t/(1\,day)$ and $E_{52}=E_{\rm iso}/(10^{52}\ \text{ergs})$,
Eq. (\ref{equation:11}) will be converted into
\begin{equation}\label{equation:12}
	\Gamma(t)=\left\{
	\begin{array}{cl}
		6.68(E_{52}/n_0)^{1/8}[t_{\rm days}/(1+z)]^{-3/8} & (k=0) \\
		4.90(E_{52}/A_*)^{1/4}[t_{\rm days}/(1+z)]^{-1/4} & (k=2) \\
	\end{array}\right.,
\end{equation}
where $n=n_0$ cm$^{-3}$ stands for a constant density with $k=0$, and the wind parameter $A_*=A/(5\times10^{11}\,g\,cm^{-1})$ is taken for the wind-like medium with $k=2$. For simplicity, a homogeneous medium around GRBs has been assumed in our calculations. Using the measured $t_b$, we then estimate the initial Lorentz factor $\Gamma$\citep{2006MNRAS.366L..13G} to be
	\begin{equation}\label{equation:10}
		\Gamma=\xi_{\rm max}\Gamma_{\rm break},
	\end{equation}
	where $\Gamma_{\rm break}=\Gamma(t_{b})$ and $\xi_{\rm max}=(t_{b}/T_{90})^{3/(8+\zeta)}$  showing the maximum ratio of the initial Lorentz factor to that at the end of the plateau \citep{2006MNRAS.366L..13G}.
	Here, $\zeta=1.5$ is defined in the differential equation
	of $dE_{\rm iso}/d\text{ln}u\propto u^\zeta$, with $u=\beta\Gamma$.

			\begin{figure}
			\centering
			\footnotesize
			\includegraphics[width=0.3\textwidth]{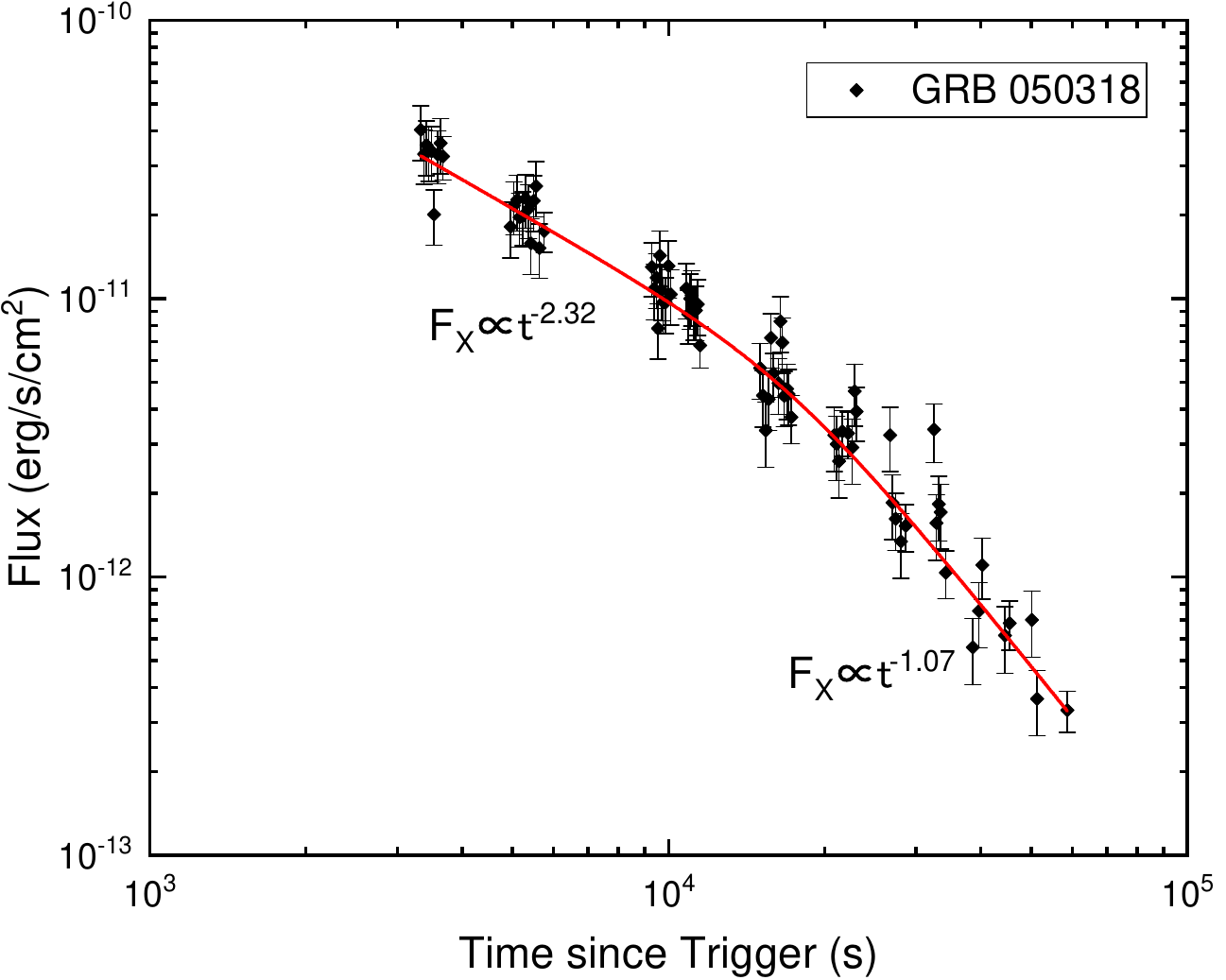}
			\includegraphics[width=0.3\textwidth]{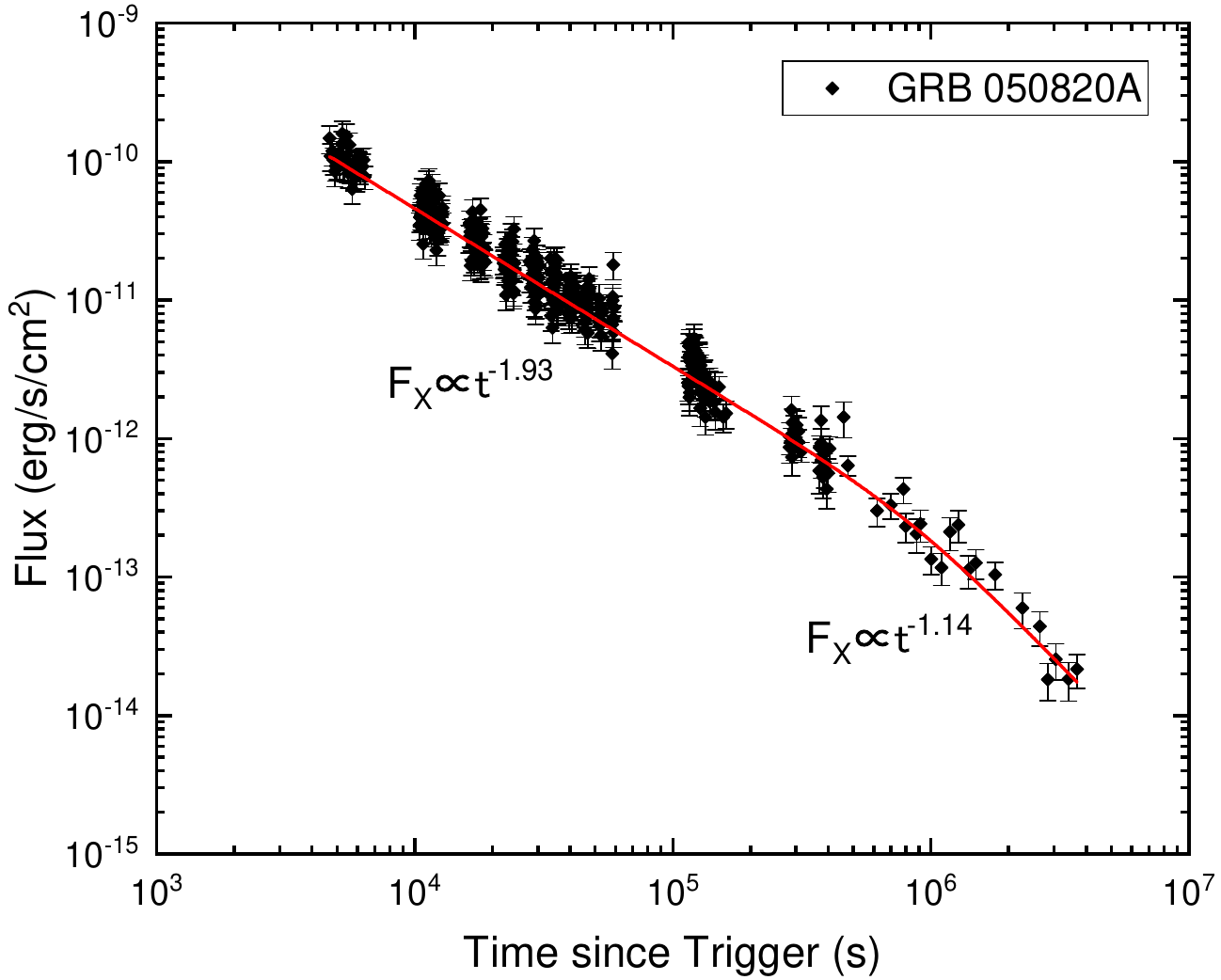}
			\includegraphics[width=0.3\textwidth]{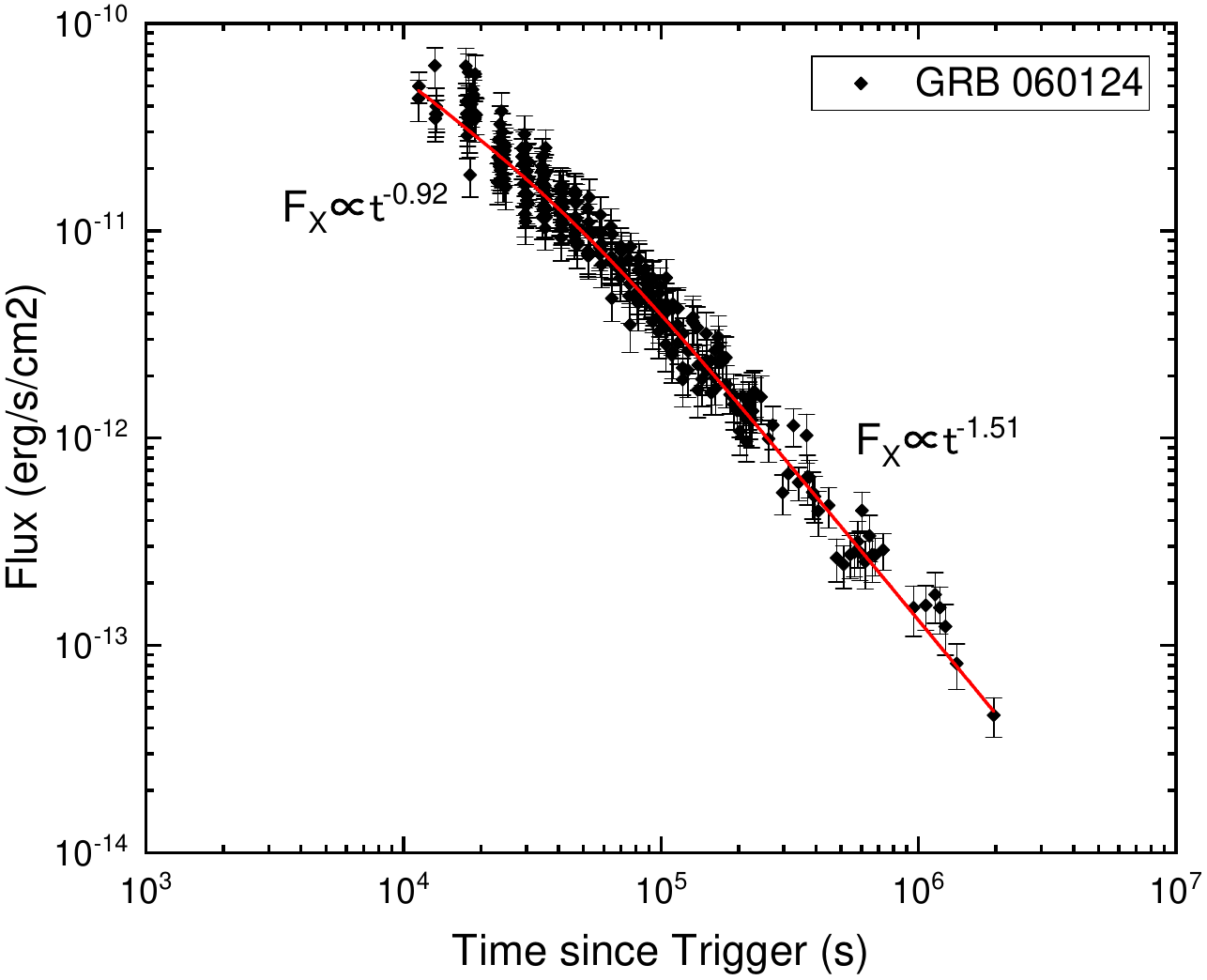}\\
			\includegraphics[width=0.3\textwidth]{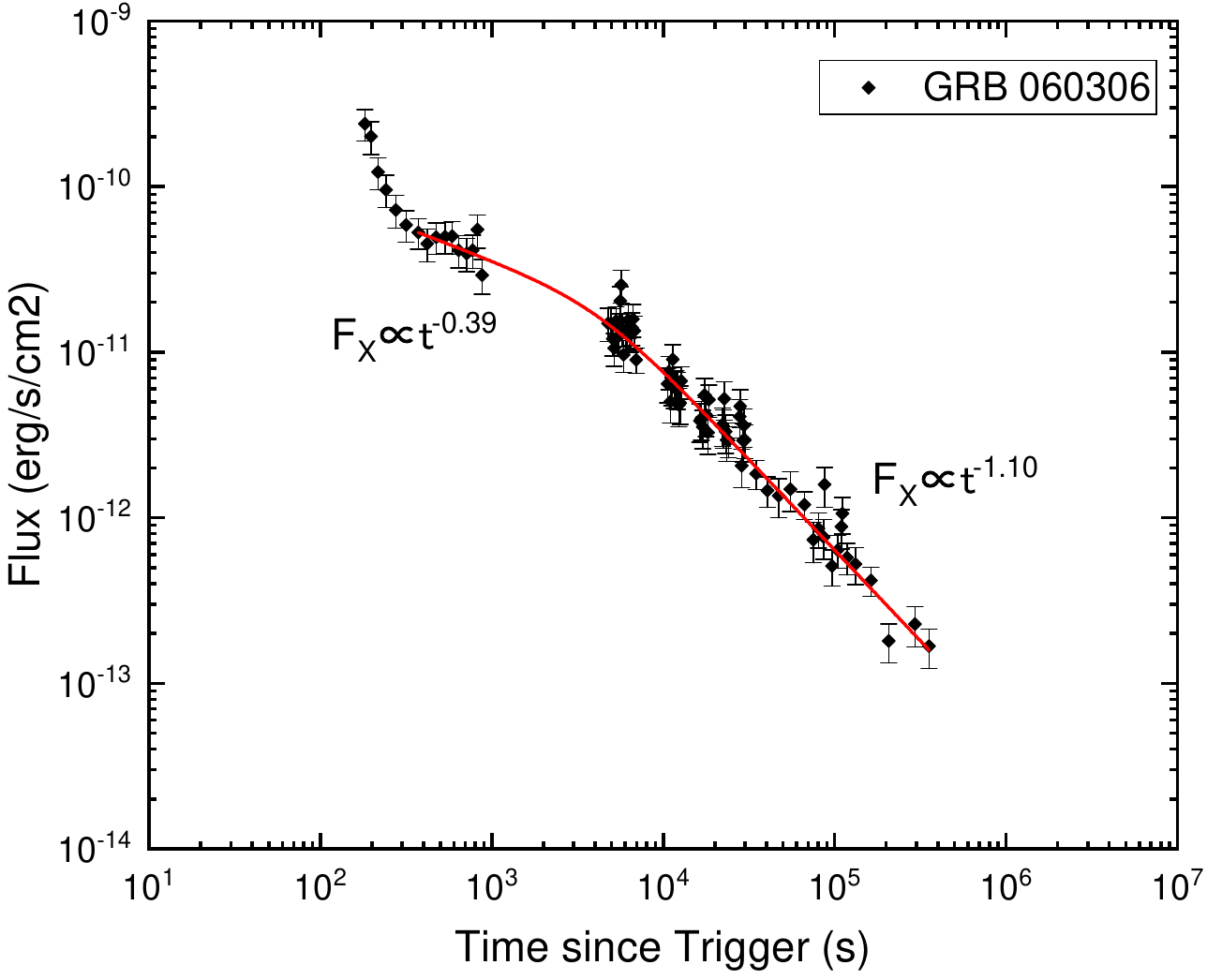}
			\includegraphics[width=0.3\textwidth]{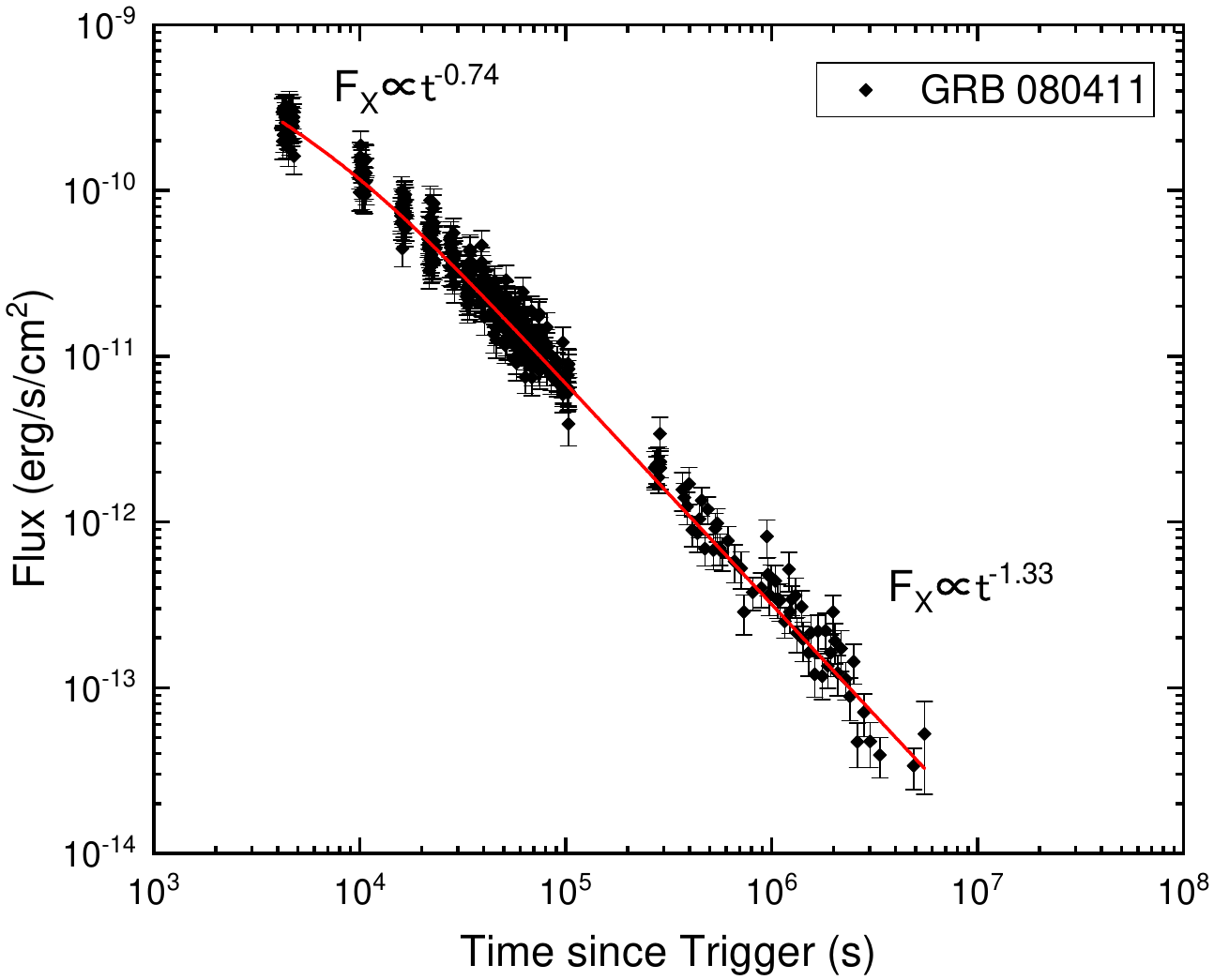}
            \includegraphics[width=0.3\textwidth]{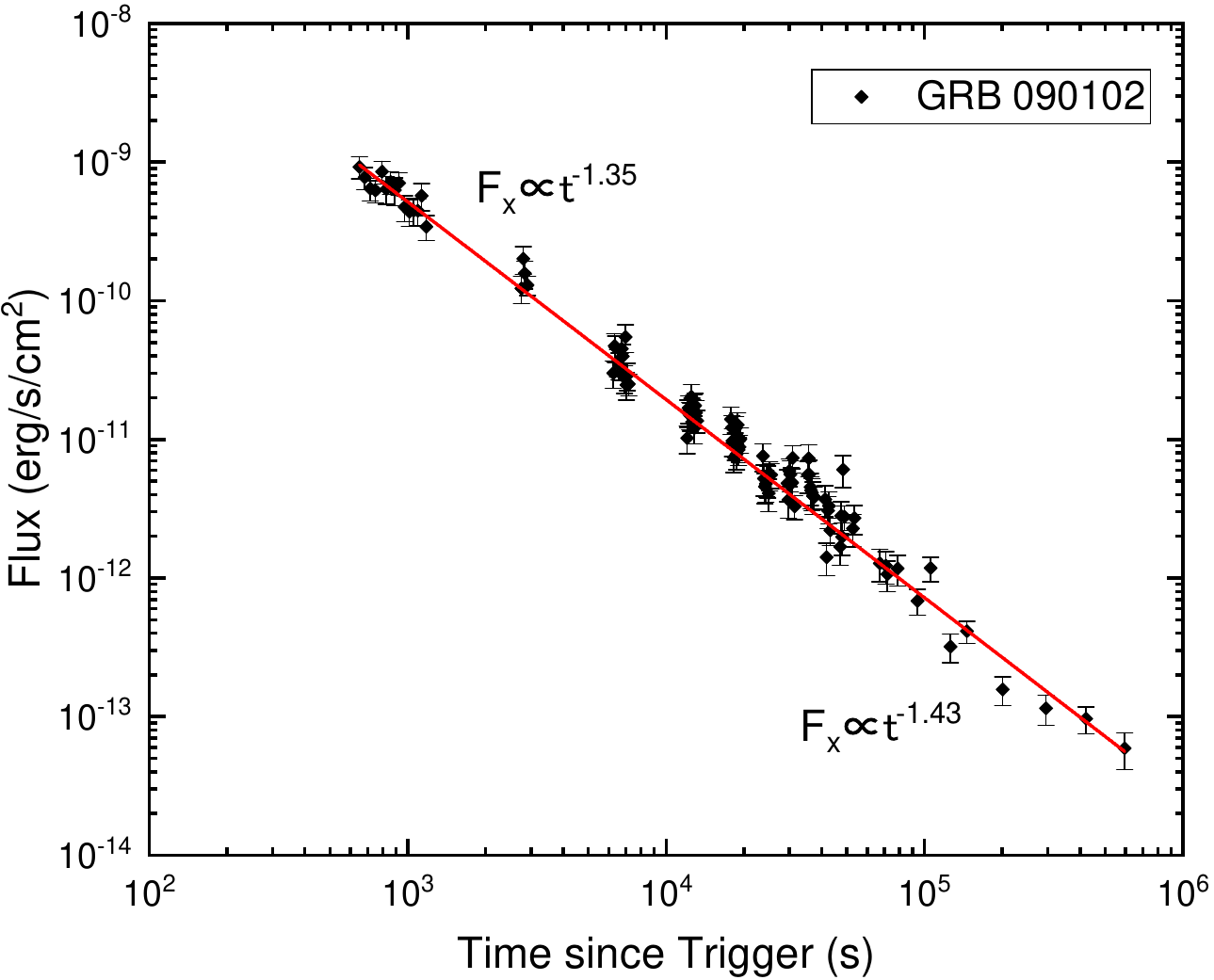}
				\caption{Six examples of the best-fitting results with Eq. (\ref{equation:9}) for X-ray light curves. Black dots correspond to the observed data.}
				\label{fig:LorentzGRB}
			\end{figure}
	\begin{figure}
			\centering
			\footnotesize
			\includegraphics[height=10cm]{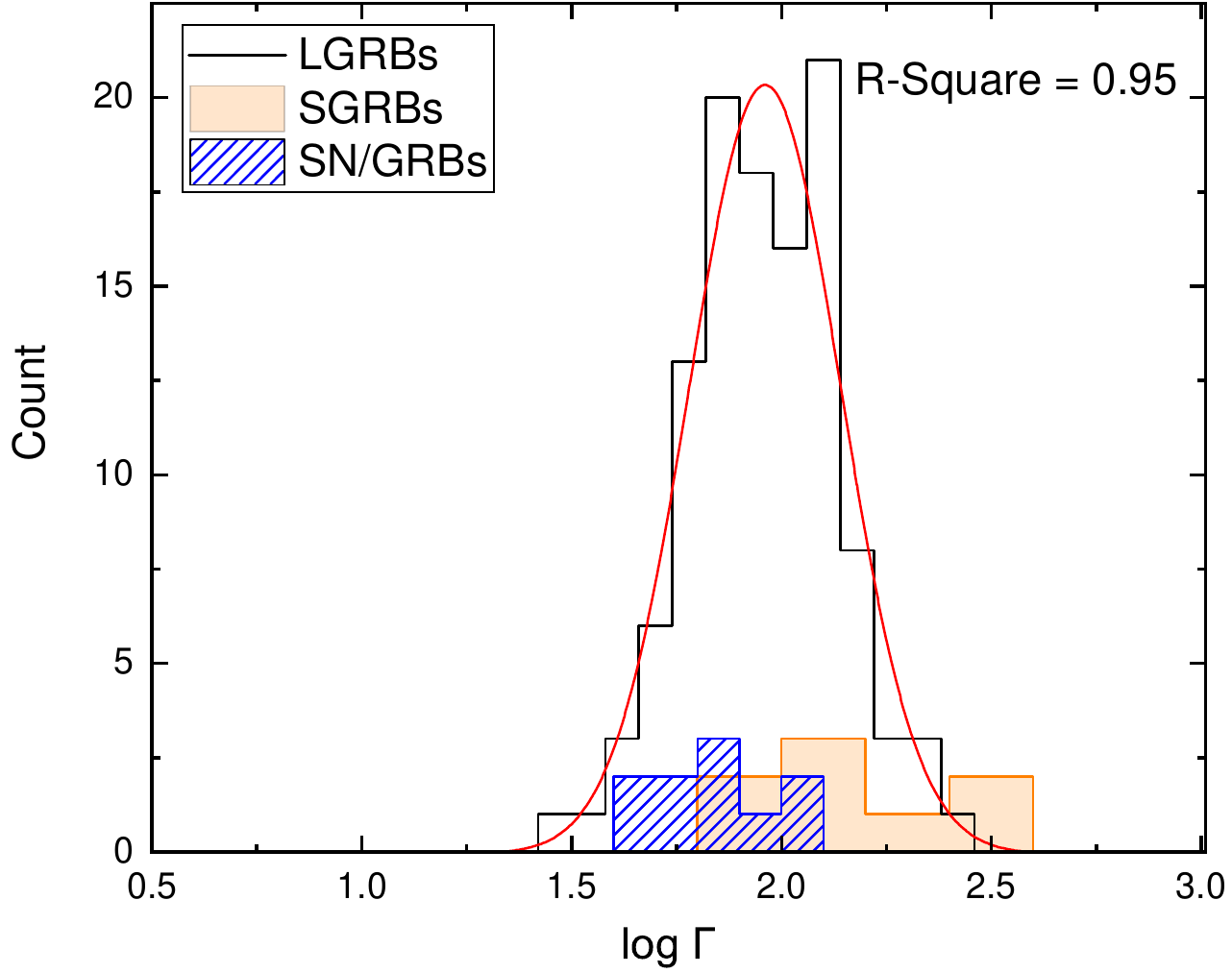}
			\caption{Histograms of the initial Lorentz factors ($\Gamma$) of different
				kinds of GRBs. All symbols have the same meanings as in Figure \ref{figs1:thetav}.}
			\label{figs3:lorentz}
		\end{figure}
Using the Eq. (\ref{equation:10}), we have measured the initial Lorentz factors of 135 GRBs with
	good $t_b$ measurements. Figure \ref{fig:LorentzGRB} shows six examples of X-ray light curves well-fitted with Eq. (\ref{equation:9}). Figure \ref{figs3:lorentz} displays the lognormal distributions of $\Gamma$,
	with a mean value of 1.96 (0.36 dex), 2.16 (0.40 dex) and 1.84 (0.30 dex)
	for LGRBs, SGRBs and SN/GRBs, respectively. This demonstrates that there is no difference in $\Gamma$ between the three classes of GRBs statistically. A larger Lorentz factor is required to explain the negligible spectral lags of SGRBs \citep{2006MNRAS.373..729Z,2006ApJ...643..266N}. However, most LGRBs have a positive spectral lags that can not be explained by the Lorentz factor only \citep{2005ApJ...627..324N,2006ChJAA...6..312Z}. For GRB pulses dominated by the angular spreading time, the duration is approximated to $T_{dur}\approx(1+z)R_{shock}/2\Gamma^2c$ that is naturally applied to infer that LGRBs are usually produced from the larger emission radii than SGRBs despite their progenitors could be distinct \citep{2007AN....328...99Z}, which indicates that the difference in emission radii would cause the inconsistency of time lags or durations between SGRBs and LGRBs. This is attributed to the fact that GRBs with long durations usually exhibit larger lags accross different energy channels.

	\section{Results}
\subsection{Parameter distributions}
Taking the observed quantities such as $E_{\rm p}^{\rm out}$, flux and fluence of GRBs with known redshifts, we follow \cite{2018PASP..130e4202Z} to calculate the out-axis quantities of $E_{\rm iso}^{\rm out}$, $L_{\rm p}^{\rm out}$ and $E_{\rm \gamma}^{\rm out}$, and then transform them into the corresponding in-axis quantities of $E_{\rm iso}^{\rm in}$, $L_{\rm p}^{\rm in}$, $E_{\rm \gamma}^{\rm in}$ and $E_{\rm p}^{\rm in}$ with Eqs. (\ref{equation:5}), (\ref{equation:7}), (\ref{equation:27}) and (\ref{equation:8}), respectively. The mean values and variances of these parameters are summarized in Table \ref{tab:tab2}, where we can see that the in-axis values are significantly larger than the corresponding out-axis ones as a whole. For instance, the mean value of $E_{\rm p}^{\rm out}$ of LGRBs and SGRBs are about one order of magnitude less than their corresponding $E_{\rm p}^{\rm in}$ mean values. However, the difference between in-axis and out-axis parameters for SN-associated GRBs (SN/GRBs) is relatively small, which may be	resulted from their narrower viewing angles. The values of $E_{\rm p}^{\rm out}$ and $E_{\rm p}^{\rm in}$ of LGRBs and SGRBs are comparable \cite[see also][]{2020ApJ...902...40Z} and they are similar to the SN/GRBs in case of out-axis but obviously harder than SN/GRBs about one order of magnitude in case of in-axis. It is notable that the similarity of peak energies between LGRBs and SGRBs has also been revealed in \cite{2018PASP..130e4202Z}. Similarly, we notice that the parameters of $E_{\rm iso}^{\rm in}$, $L_{\rm p}^{\rm in}$ and$E_{\rm \gamma}^{\rm in}$ are roughly one order of magnitude larger than the corresponding out-axis parameters for both LGRBs and SGRBs, in	contrast to SN/GRBs whose difference is relatively small. In total, LGRBs in our sample have the largest energy release and an intermediate hardness ratio in both out-axis and in-axis parameters.
\begin{table}
	\centering
	\caption{Statistical properties of different types of out-axis GRBs }
	\label{tab:tab2}
		\begin{tabular}{cccc}
			\hline
			& Type & Mean  & Sample size \\ \hline
			\multirow{3}{*}{\shortstack{$\rm log\textit{T}_{\rm 90}$\\(s)}}                      & $SGRB$ &  $ -0.02\pm0.32   $     &      8          \\
			& $LGRB$ &   $1.58\pm0.50 $      &      128         \\ 
			& $SN/GRBs$  &  $1.21\pm0.78   $    &      12          \\ \hline
			\multirow{3}{*}{$\rm log\textit{z}$}                                             & $SGRB$ &  $-0.08\pm0.32 $       &      8          \\ 
			& $LGRB$ &   $0.30\pm0.27  $      &      128         \\ 
			& $SN/GRBs$  &   $-0.31\pm0.38 $       &      12          \\ \hline
			\multirow{3}{*}{\shortstack{\textbf{$E_{\rm p}^{\rm out}$}\\($\rm keV$)}}               & $SGRB$ &   $158.9\pm43.2 $       &      8          \\ 
			& $LGRB$ &  $ 150.9\pm31.8 $       &      128         \\ 
			& $SN/GRBs$  &   $124.0\pm20.7     $   &      12          \\ \hline
			\multirow{3}{*}{\shortstack{\textbf{$E_{\rm p}^{\rm in}$}\\$(\rm keV)$}}                & $SGRB$ &   $1706.3\pm515.1 $      &      8           \\ 
			& $LGRB$ &   $1241.9\pm222.1 $      &      114         \\ 
			& $SN/GRBs$  &  $ 194.9\pm22.3 $       &      10          \\ \hline
			\multirow{3}{*}{\shortstack{$E_{\rm iso}^{\rm out}$\\$(10^{52} \rm erg)$} }    & $SGRB$ &   $0.24\pm0.02 $      &      8          \\ 
			& $LGRB$ &  $ 18.2\pm0.70$       &      128         \\ 
			& $SN/GRBs$  & $12.5\pm0.16 $      &      12          \\ \hline
			\multirow{3}{*}{\shortstack{$E_{\rm iso}^{\rm in}$\\$(10^{52} \rm erg)$}}      & $SGRB$ &  $ 2.44\pm0.26$       &      8           \\ 
			& $LGRB$ &   $147.5\pm4.14 $      &      114         \\ 
			& $SN/GRBs$  & $  20.4\pm0.25  $     &      10          \\ \hline
			\multirow{3}{*}{\shortstack{$L_{\rm p}^{\rm out}$\\$(10^{51} \rm erg\ \rm s^{-1})$}}& $SGRB$ &   $5.14\pm0.51$       &      8          \\ 
			& $LGRB$ & $  46.3\pm4.11 $      &      128         \\ 
			& $SN/GRBs$  &  $ 7.66\pm0.26  $     &      12          \\ \hline
			\multirow{3}{*}{\shortstack{$L_{\rm p}^{\rm in}$\\$(10^{51} \rm erg\ \rm s^{-1})$}}& $SGRB$ &$   61.6\pm6.44 $      &      8           \\ 
			& $LGRB$ &  $ 426.5\pm31.6$       &      114         \\ 
			& $SN/GRBs$  &  $ 15.4\pm0.48     $  &      10          \\ \hline
			\multirow{3}{*}{\shortstack{$E_{\rm \gamma}^{\rm out}$\\$(10^{50} \rm erg)$}}  & $SGRB$ & $  1.26\pm  0.12 $      &      8          \\ 
			& $LGRB$ &   $55.9\pm1.65  $     &      128         \\ 
			& $SN/GRBs$  &   $5.58\pm0.09  $     &      11          \\ \hline
			\multirow{3}{*}{\shortstack{$E_{\rm \gamma}^{\rm in}$\\$(10^{50} \rm erg)$}}   & $SGRB$ &   $15.53\pm1.80    $   &      8          \\ 
			& $LGRB$ &   $734.7\pm23.2 $      &      114         \\ 
			& $SN/GRBs$  &  $ 9.97\pm0.14 $      &      10          \\ \hline\hline
		\end{tabular}
	\end{table}
	
	\subsection{Spectrum-energy relations with instrinsic dispersions }
	\begin{figure*}
		\centering
		\includegraphics[width=0.49\textwidth]{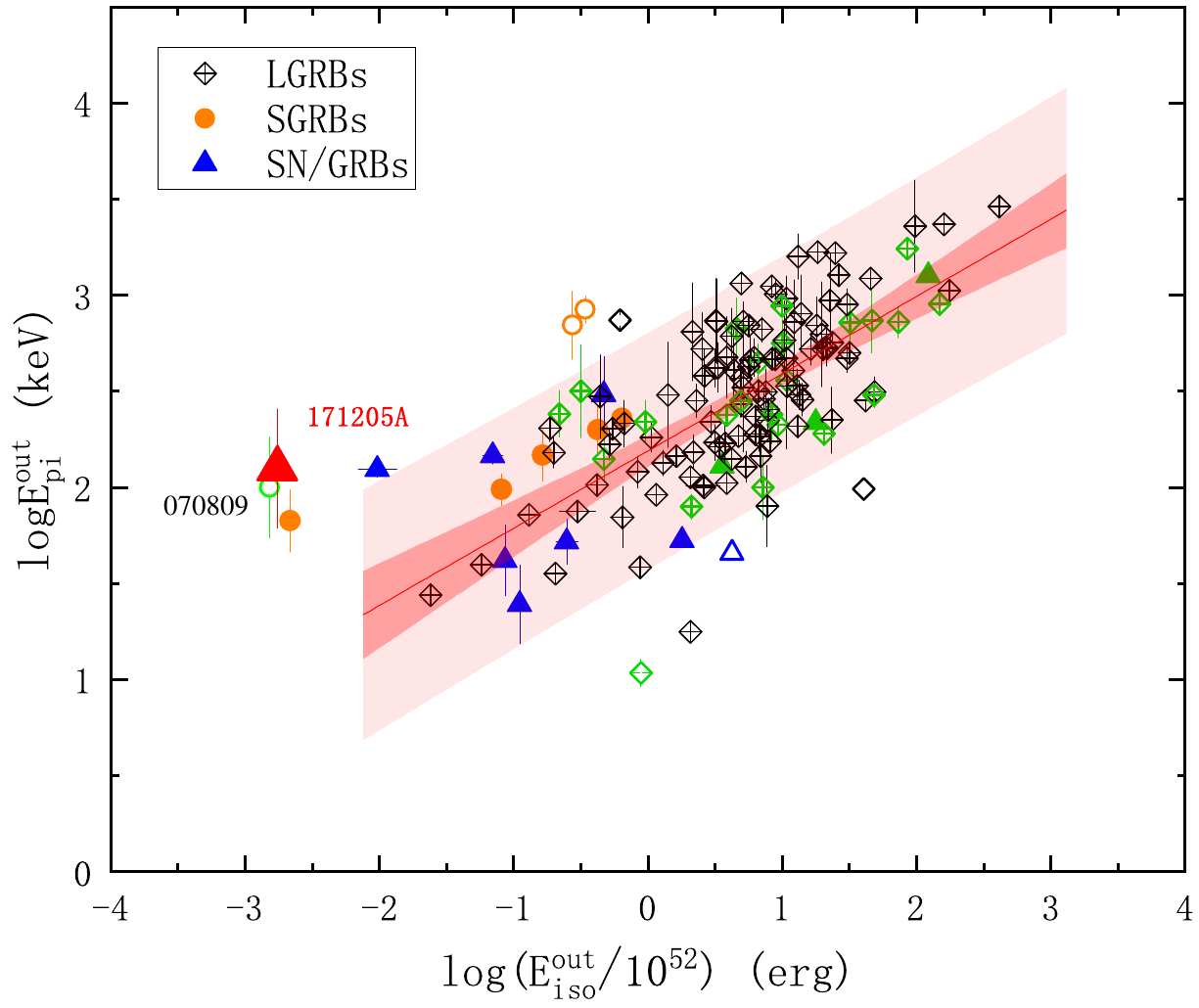}
		\includegraphics[width=0.49\textwidth]{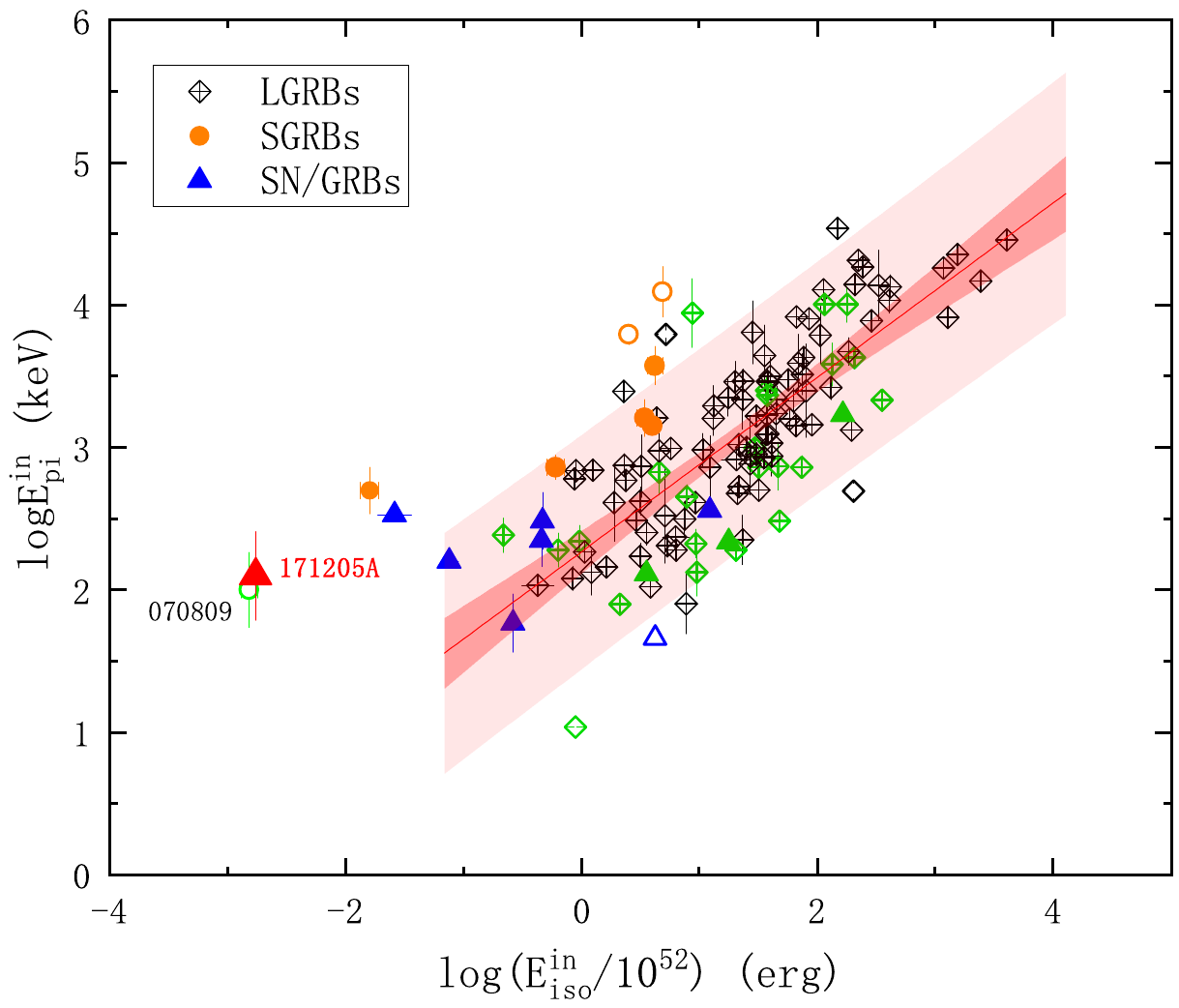}
			\caption{The $E_{\rm pi}-E_{\rm{iso}}$ relations of out-axis (left panel)  and in-axis (right panel) GRBs. The solid lines denote the best fits to data. The heavy and light shade regions represent confidence and prediction regions in $2\sigma$ level, respectively. Note that the empty symbols indicate that these bursts deviating from the $E_{\rm pi}-E_{\rm{iso}}$ relation in both in- and out-axis cases. The green diamonds stand for those bursts with well-constrained $\theta_{\rm v}$ in Table \ref{tab:sample}.
			}
			\label{fig:SE_A}
		\end{figure*}
		\begin{figure*}
			\centering
			\footnotesize
			\includegraphics[width=0.49\textwidth]{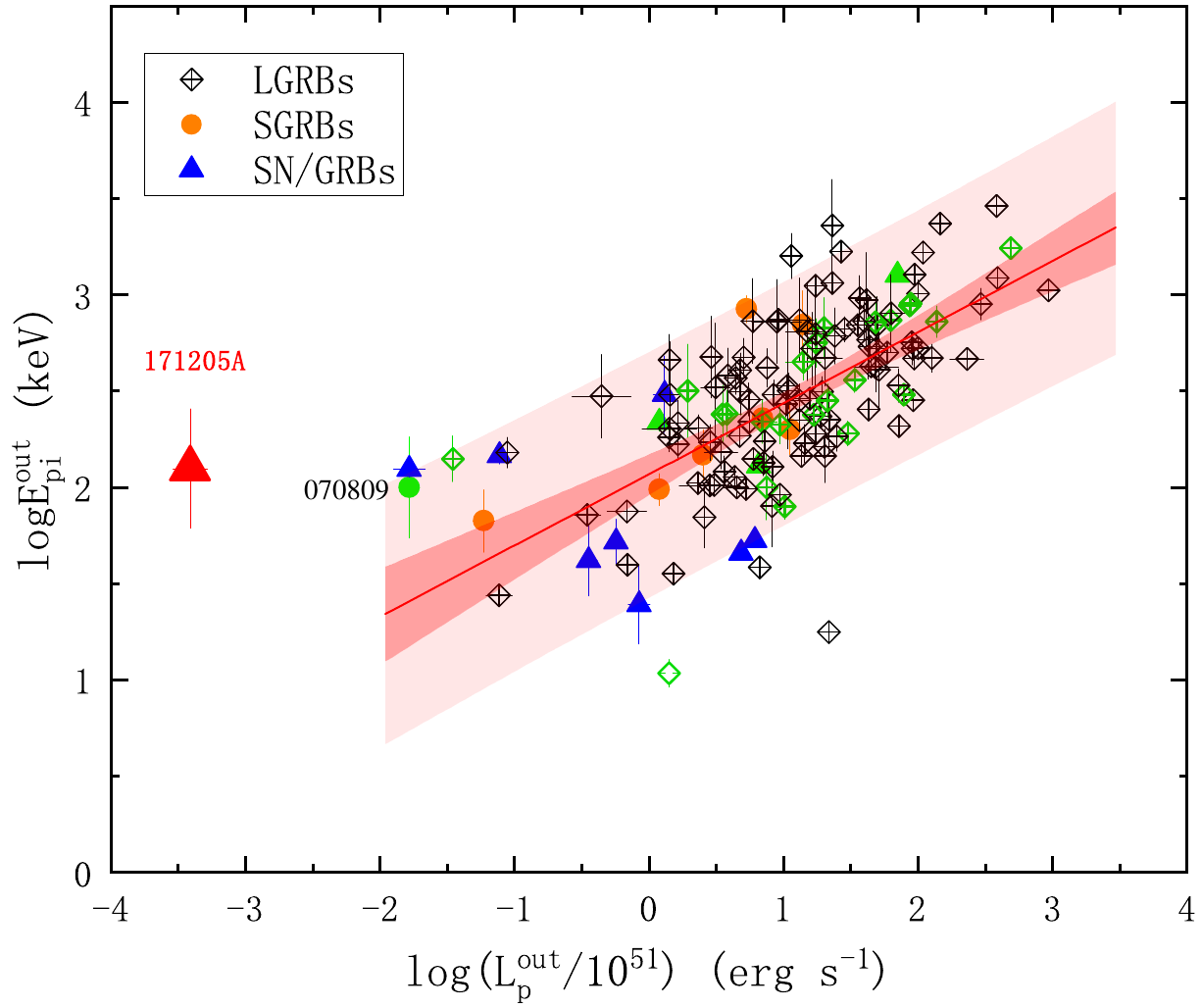}
			\includegraphics[width=0.49\textwidth]{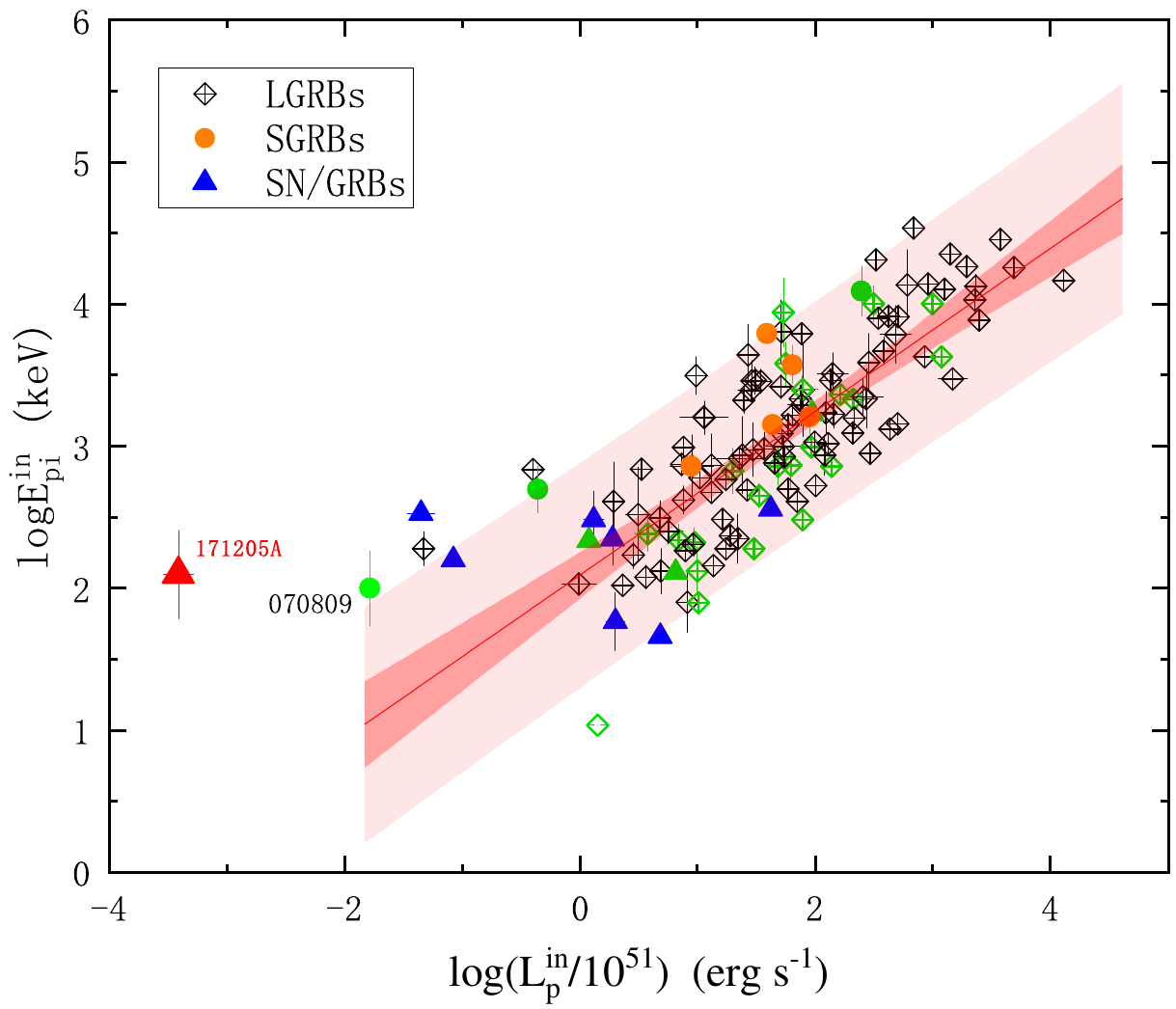}
				\caption{The $E_{\rm pi}-L_{\rm{p}}$ relations of out-axis (left panel) and in-axis (right panel) LGRBs. All symbols are the same as in Figure \ref{fig:SE_A}.}
				\label{fig:SE_Y}
			\end{figure*}
			
			\begin{figure*}
				\centering
				\includegraphics[width=0.49\textwidth]{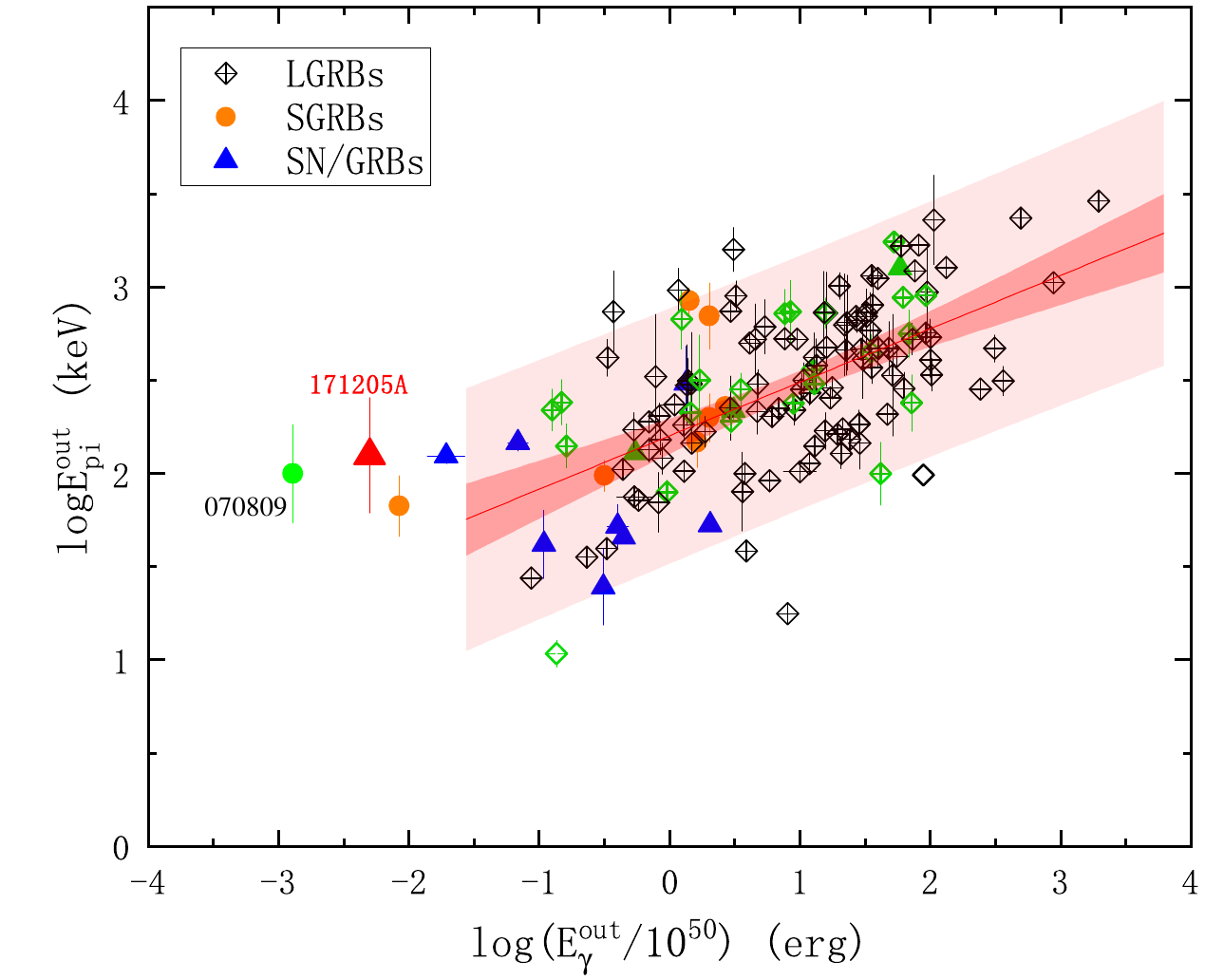}
				\includegraphics[width=0.49\textwidth]{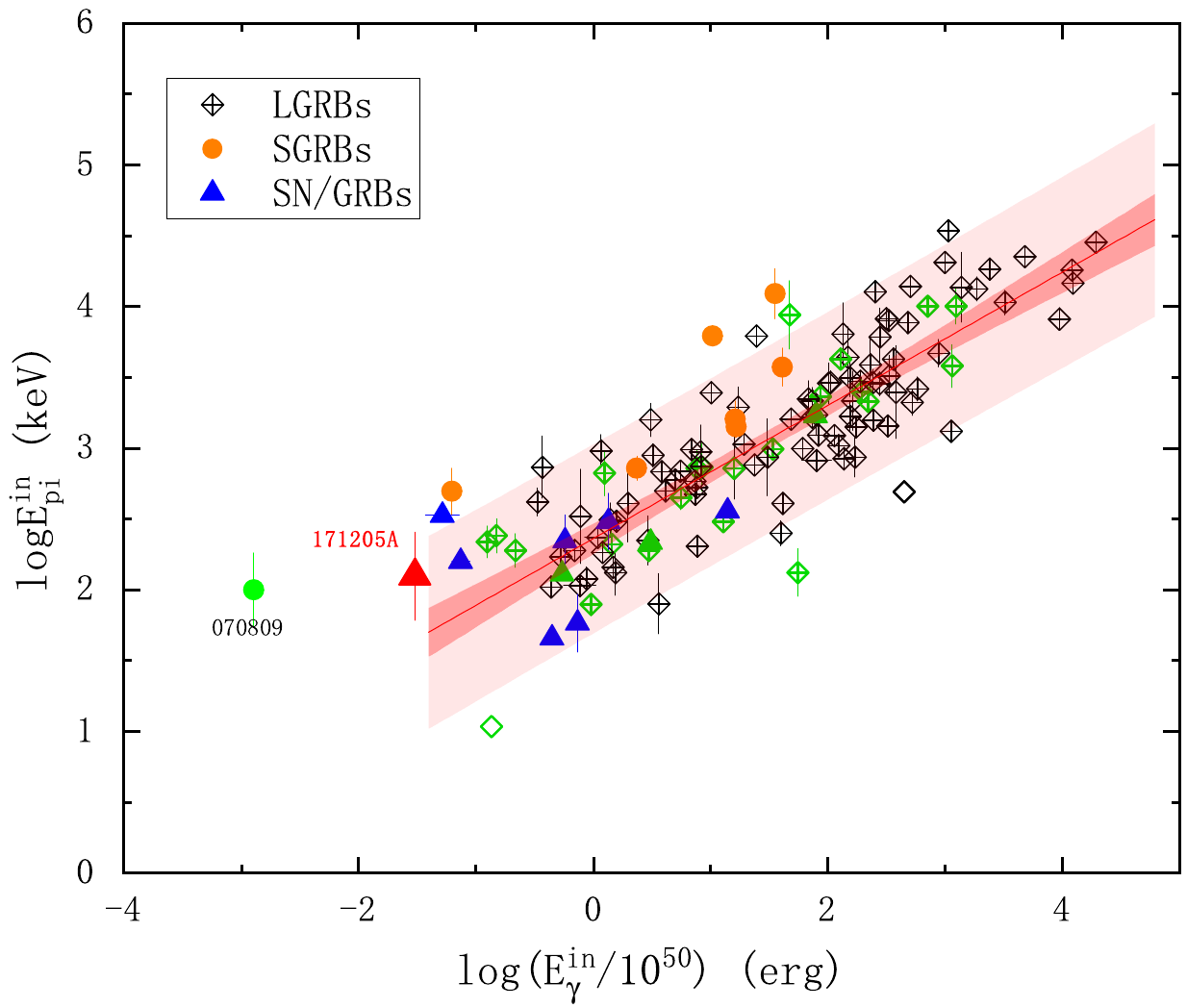}
					\caption{The $E_{\rm pi}-E_{\rm{\gamma}}$ relations of out-axis (left panel) and in-axis (right panel) LGRBs. All symbols are the same as in Figure \ref{fig:SE_A}.}
					\label{fig:SE_G}
				\end{figure*}
				\begin{figure*}
					\centering
					\includegraphics[width=1.0\textwidth]{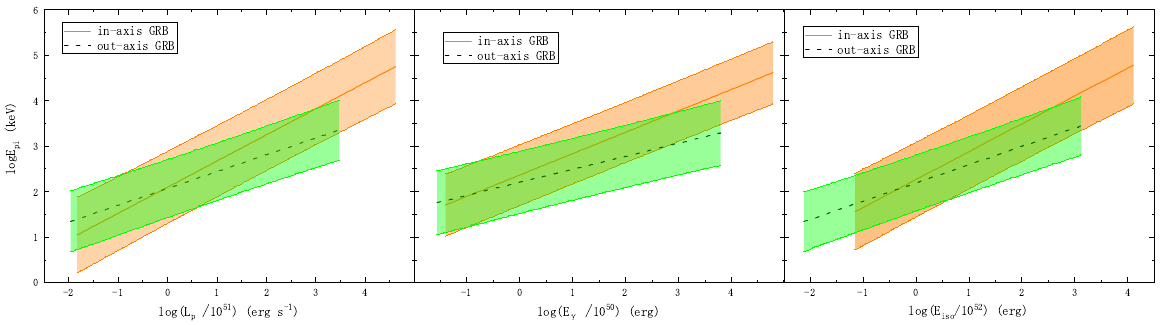}
					\caption{The out-axis (dashed line) and in-axis (solid line) correlations of $E_{\rm pi}$ vs. $L_{\rm p}$ extracted from Fig. 7 (left panel),  $E_{\rm pi}$ vs. $E_{\rm \gamma}$ extracted from Fig. 8 (middle panel) and $E_{\rm pi}$ vs. $E_{\rm iso}$ extracted from Fig. 6 (right panel). The shaded regions show the 3$\sigma$ confidence ranges of these energy correlations.}
					\label{fig:8}
				\end{figure*}
				
We now utilize the rest-frame peak energies of $E_{\rm pi}^{\rm in}$ and $E_{\rm pi}^{\rm out}$ to plot against $E_{\rm iso}^{\rm in}$ and $E_{\rm iso}^{\rm out}$ correspondingly for LGRBs, SGRBs and SN/GRBs in Figure \ref{fig:SE_A}, where the best fitting $E_{\rm pi}-E_{\rm{iso}}$ relations \citep{2002A&A...390...81A} of $E_{\rm pi}=C_A(E_{\rm iso}/10^{52}\ erg)^{\eta_{\rm A}}$ for LGRBs except GRB		171205A can be written as
\begin{equation}
					E_{\rm pi}^{\rm out}=154.88^{+13.63}_{-14.94}\left(\frac{E_{\rm iso}^{\rm out}}{10^{52}erg}\right)^{0.40\pm0.04 } (\textrm{keV}), \label{equation:13}
				\end{equation}
				and
				\begin{equation}
					E_{\rm pi}^{\rm in}=186.21^{+31.33}_{-37.66}\left(\frac{E_{\rm iso}^{\rm in}}{10^{52}erg}\right)^{0.61\pm0.05 } (\textrm{keV}), \label{equation:14}
				\end{equation}
				with the Spearman's rank correlation coefficient being 0.63 and
				0.78, respectively. We see that the in-axis $E_{\rm pi}-E_{\rm{iso}}$ correlation is
				steeper than the out-axis one. The out-axis power-law index of
				$\eta_A^{\rm out}\approx0.40$ is consistent with $0.35\pm0.06$ in \cite{2003ChJAS...3..455A} and $0.40\pm0.05$ in \cite{2004ApJ...616..331G}. Nevertheless, our in-axis power-law index of
				$\eta_A^{\rm in}\approx0.61$ coincides with $0.52\pm0.06$ in \cite{2002A&A...390...81A}, $0.57\pm0.02$ in both \cite{2006MNRAS.372..233A} and \cite{2006A&A...450..471N}. Notably, \cite{2018PASP..130e4202Z} found that LGRBs and SGRBs uncorrected for the out-axis effect possess the identical slope of $\eta\approx0.35$, which is roughly in accord with $\eta_A^{\rm out}$. Moreover,  GRBs with EE were found to have a larger slope of $0.45\pm0.05$ for the E-I type and a smaller slope of $0.36\pm0.04$ for the E-II type. Most slopes derived for distinct GRB samples deviate from the
				theoretical value of 0.5 slightly \citep[][X23]{2004ApJ...616..331G,2005ApJ...627....1F}, which may be biased by several effects including sensitivity of detectors, sample selection, energy band, viewing angle, mixture of varieties and so on. 
			
			In the same
				way, we have plotted the out-axis and in-axis \textbf{$E_{\rm pi}-L_{\rm{p}}$}
				spectrum-energy relations \citep{2004ApJ...609..935Y} of $E_{\rm pi}=C_Y(L_{\rm p}/10^{51}\
				erg)^{\eta_Y}$ for LGRBs in Figure \ref{fig:SE_Y}. The best fit
				results are
				\begin{equation}
					E_{\rm pi}^{\rm out}=117.49^{+12.78}_{-14.34}\left(\frac{L_{\rm p}^{\rm out}}{10^{51}erg\ s^{-1}}\right)^{0.37\pm0.04 } (\textrm{keV}),\label{equation:15}
				\end{equation}
				and
				\begin{equation}
					E_{\rm pi}^{\rm in}=123.03^{+20.70}_{-24.88}\left(\frac{L_{\rm p}^{i\rm n}}{10^{51}erg\ s^{-1}}\right)^{0.57\pm0.04 } (\textrm{keV}) ,\label{equation:16}
				\end{equation}
				with the Spearman's rank correlation coefficient being 0.66 and
				0.80, respectively. Again, the out-axis \textbf{$E_{\rm pi}-L_{\rm{p}}$} relation is also
				flatter than the in-axis one. Note that the out-axis power-law
				index is consistent with our previous results for both regular \citep{2004ApJ...609..935Y,2018PASP..130e4202Z} and
				EE GRBs \citep{2020RAA....20..201Z}. Similarly, for LGRBs with
				measured half-opening jet angle, the $E_{\rm pi}-E_{\rm{\gamma}}$ relation \citep{2004ApJ...616..331G} of
				$E_{\rm pi}=C_G(E_{\rm \gamma}/10^{50}\ erg)^{\eta_{\rm G}}$ is plotted in
				Figure \ref{fig:SE_G}, and the best fit result is
				\begin{equation}
					E_{\rm pi}^{\rm out}=158.49^{+17.24}_{-19.34}\left(\frac{E_{\rm \gamma}^{\rm out }}{10^{50}erg}\right)^{0.29\pm0.04 } (\textrm{keV}), \label{equation:17}
				\end{equation}
				and
				\begin{equation}
					E_{\rm pi}^{\rm in}=229.09^{+24.91}_{-27.95}\left(\frac{E_{\rm \gamma}^{\rm in}}{10^{50}erg}\right)^{0.47\pm0.03 } (\textrm{keV}),  \label{equation:18}
				\end{equation}
with the Spearman's rank correlation coefficient being 0.53 and 0.86 for the out- and in-axis cases, respectively. We see that the in-axis \textbf{$E_{\rm pi}-E_{\rm{\gamma}}$} relation is much steeper than the out-axis one. Interestingly, the beaming-corrected spectrum-energy relation of	the out-axis case has almost the same power-law index as that of regular bursts and GRBs with EE presented in our previous studies \citep{2018PASP..130e4202Z,2020RAA....20..201Z}.
				
It is worthy of attention that we have followed \cite{Kumar2023} to adopt the Maximum Likelihood Estimation (MLE) method \citep{2005physics..11182D} to obtain the above Eqs. (\ref{equation:13}-\ref{equation:18}) together with intrinsic scatters ($\sigma_s$). For this purpose, we rewrite the Amati relation as an example in logarithmic scale as
\begin{equation}
	\log E_{\rm pi} = q_{\rm A} + \eta_{\rm A} \log E_{{\rm iso},52},
	\label{eq:log_amati}
\end{equation}
where $E_{{\rm iso},52} \equiv E_{\rm iso}/10^{52}$~erg, and $q_{\rm A}\equiv logC_A$. Note that $C_A$ and $\eta_{A}$ are two fitted parameters as shown in Eqs. (\ref{equation:13}-\ref{equation:14}). Defining the variables:
\begin{equation}
	y \equiv \log\left[\frac{E_{\rm pi}}{1~{\rm keV}}\right] \ \rm{and} \qquad
	x \equiv \log\left[\frac{E_{\rm iso,52}}{1~{\rm erg}}\right],
	\label{eq:xy_def}
\end{equation}
one can express Eq. (\ref{eq:log_amati}) as 
\begin{equation}
	y = \eta_{\rm A} x + q_{\rm A}.
	\label{eq:linear_model}
\end{equation}
The uncertainties of $y$ and $x$ are given by the error propagation to be
\begin{equation}
	\sigma_{y} = \frac{1}{\ln 10} \left( \frac{\sigma_{E_{\rm pi}}}{E_{\rm pi}} \right), \quad
	\sigma_{x} = \frac{1}{\ln 10} \left( \frac{\sigma_{E_{\rm iso,52}}}{E_{\rm iso,52}} \right).
	\label{eq:errors}
\end{equation}
Assuming $y$ obeys a Gaussian distribution, we can specify the likelihood function $\mathcal{L}$ as
\begin{equation}
	-2\ln\mathcal{L} = \sum_{i=1}^{N} \ln(2\pi\sigma_{i}^{2}) + \sum_{i=1}^{N} \frac{\left[y_i - (\eta_{\rm A} x_i + q_{\rm A})\right]^{2}}{\sigma_{Ai}^{2}},
	\label{eq:likelihood}
\end{equation}
where the total variance $\sigma_{Ai}^2$ for the $i$-th GRB is determined by
\begin{equation}
	\sigma_{Ai}^2 = \sigma_{y_i}^2 + \eta_{\rm A}^2 \sigma_{x_i}^2 + \sigma_{sA}^2,
	\label{eq:total_variance}
\end{equation}
in which $\sigma_{y_i}$ and $\sigma_{x_i}$ represent the measurement uncertainties, while $\sigma_{sA}$ indicates the intrinsic scatter or tightness of the $E_{\rm pi}$-$E_{\rm iso}$ relation. Similarly, we can derive the total variances of $\sigma_{Yi}^2 = \sigma_{y_i}^2 + \eta_{\rm Y}^2 \sigma_{x_i}^2 + \sigma_{sY}^2$ with $x \equiv \log[L_{\rm p,51}/(\rm erg\ s^{-1})]$ for the Yonetoku relation and $\sigma_{Gi}^2 = \sigma_{y_i}^2 + \eta_{\rm G}^2 \sigma_{x_i}^2 + \sigma_{sG}^2$ with $x \equiv \log[E_{\rm \gamma,50}/{\rm erg}]$ for the Ghirlanda relation, respectively.
The best fitted parameters including three intrinsic dispersions ($\sigma_{sA}$, $\sigma_{sY}$ and $\sigma_{sG}$) of the out/in-axis	spectrum-energy correlations concerning LGRBs are summarized in Table \ref{tab:tab3}. Both SN/GRBs and SGRBs have not been fitted due to the limited numbers. They are specially
				marked for comparison in Figs. \ref{fig:SE_A}--\ref{fig:SE_G}. It can be found that all the in-axis
				spectrum-energy relations are ubiquitously steeper than the
				corresponding out-axis ones, which is in good agreement with the
				recent results of other groups \citep[][X23]{2021MNRAS.501.5723F}.
				Interestingly, this phenomenon can be satisfactorily explained in	the framework of the synchrotron radiation mechanism by considering the variation of the Lorentz factor and the 		corresponding relativistic boosting effect (X23). The low-luminosity GRB 171205A associated		with the type Ic SN 2017iuk \citep{2017ATel11038....1D} is an obvious outlier of
				the \textbf{$E_{\rm pi}-E_{\rm{iso}}$} and the \textbf{$E_{\rm pi}-L_{\rm{p}}$} relations. Based on multiple-wavelength
				observations, \cite{2021ApJ...907...60M} concluded that GRB 171205A should have been viewed sideways.
				
The out-axis energy relations of \textbf{$E_{\rm pi}-E_{\rm{iso}}$}, \textbf{$E_{\rm pi}-L_{\rm{p}}$} and \textbf{$E_{\rm pi}-E_{\rm{\gamma}}$} are compared with the corresponding in-aixs ones in Figure \ref{fig:8}, where we can find that the out-axis energy relations are flatter than the in-axis relations since the out-axis energies are significantly smaller than the in-axis values. Our results of the \textbf{$E_{\rm pi}-E_{\rm{iso}}$} relation are consistent with \cite{2019MNRAS.487.4884I} and X23. The \textbf{$E_{\rm pi}-L_{\rm{p}}$} relations in both cases are in good agreement with X23. In addition, we find that the \textbf{$E_{\rm pi}-E_{\rm{\gamma}}$} relation of in-axis GRBs is relatively steeper than the out-axis relation as well. This demonstrates that the in-jet energy correlations will be also steeper provided that the jets are structured and viewed out-axis. Furthermore, we find that the outliers in the in- and out-axis cases are somewhat different, especially for the $E_{\rm pi}$-$E_{\rm iso}$ correlation. 
		\begin{table}
			\centering
			\caption{The best fitted parameters of 128 out-axis GRBs}
			\label{tab:tab3}
			\begin{tabular}{c|c c c c c c}
				\hline
				& Correlation & $\eta$ & $q$ & $C$/keV & $\sigma_s$ & $\chi^2/$dof\\
				\hline
				\multirow{3}{*}{Out-axis LGRBs }
				& $E_{\rm pi}-E_{\rm iso}$ & 0.40$\pm$0.04 & 2.19$\pm$0.04 & $154.88^{+13.63}_{-14.94}$ & 0.31$\pm$0.02 & 125.01/128\\
				& $E_{\rm pi}-L_{\rm p}$ & 0.37$\pm$0.04 & 2.07$\pm$0.05 & $117.49^{+12.78}_{-14.34}$ & 0.32$\pm$0.02 & 125.72/128\\
				& $E_{\rm pi}-E_{\gamma}$ & 0.29$\pm$0.04 & 2.20$\pm$0.05 & $158.49^{+17.24}_{-19.34}$ & 0.34$\pm$0.02 & 125.71/128\\
				\hline
				\multirow{3}{*}{In-axis LGRBs }
				& $E_{\rm pi}-E_{\rm iso}$ & 0.61$\pm$0.05 & 2.27$\pm$0.08 & $186.21^{+31.33}_{-37.66}$ & 0.41$\pm$0.03 & 112.83/114\\
				& $E_{\rm pi}-L_{\rm p}$ & 0.57$\pm$0.04 & 2.09$\pm$0.08 & $123.03^{+20.70}_{-24.88}$ & 0.39$\pm$0.03 & 112.80/114\\
				& $E_{\rm pi}-E_{\gamma}$ & 0.47$\pm$0.03 & 2.36$\pm$0.05 & $229.09^{+24.91}_{-27.95}$ & 0.33$\pm$0.02 & 113.50/114\\
				\hline
			\end{tabular}
		\end{table}		

				\subsection{Hubble Diagram}
				GRB Hubble diagram can be used to explore the expansion history of
				the universe at high redshift. To construct a meaningful GRB
				Hubble diagram, some empirical spectrum-energy relations are
				usually applied to make GRBs as standard candles as far as
				possible \citep[e.g.][]{2007ApJ...660...16S,2008MNRAS.391..577A,2016A&AT...29..193A,2018pgrb.book.....Z}. As illustrated above, the
				spectrum-energy relations themselves are different for out-axis
				and in-axis cases. So, we should be careful in drawing the GRB
				Hubble diagram by using them. The luminosity distance of a GRB is
			\begin{equation}
				D_{\rm L}=\frac{{(1+z)}c}{H_{0}}\int^{z}_{0}[\Omega_{m}(1+z^{'})^{3}+\Omega_{\Lambda}]^{-0.5}{dz^{'}}.
			\end{equation}
				Taking into account the calibrated spectrum-energy relations of
				Equations (1)-(6), we can derive the distance moduli as
				\begin{equation}
					\textrm{(I)} \ \ \ \ \ \ \mu_{GRB}=25+\frac{5}{2}\left[\frac{log E_{\rm pi}-q_{\rm A}}{\eta_{A}}-log\left(\frac{4\pi S_{\rm bol}}{1+z}\right)+52\right] ,\label{equation:23}
				\end{equation}			
				\begin{figure*}
					\centering
					\footnotesize
					\includegraphics[width=0.49\textwidth]{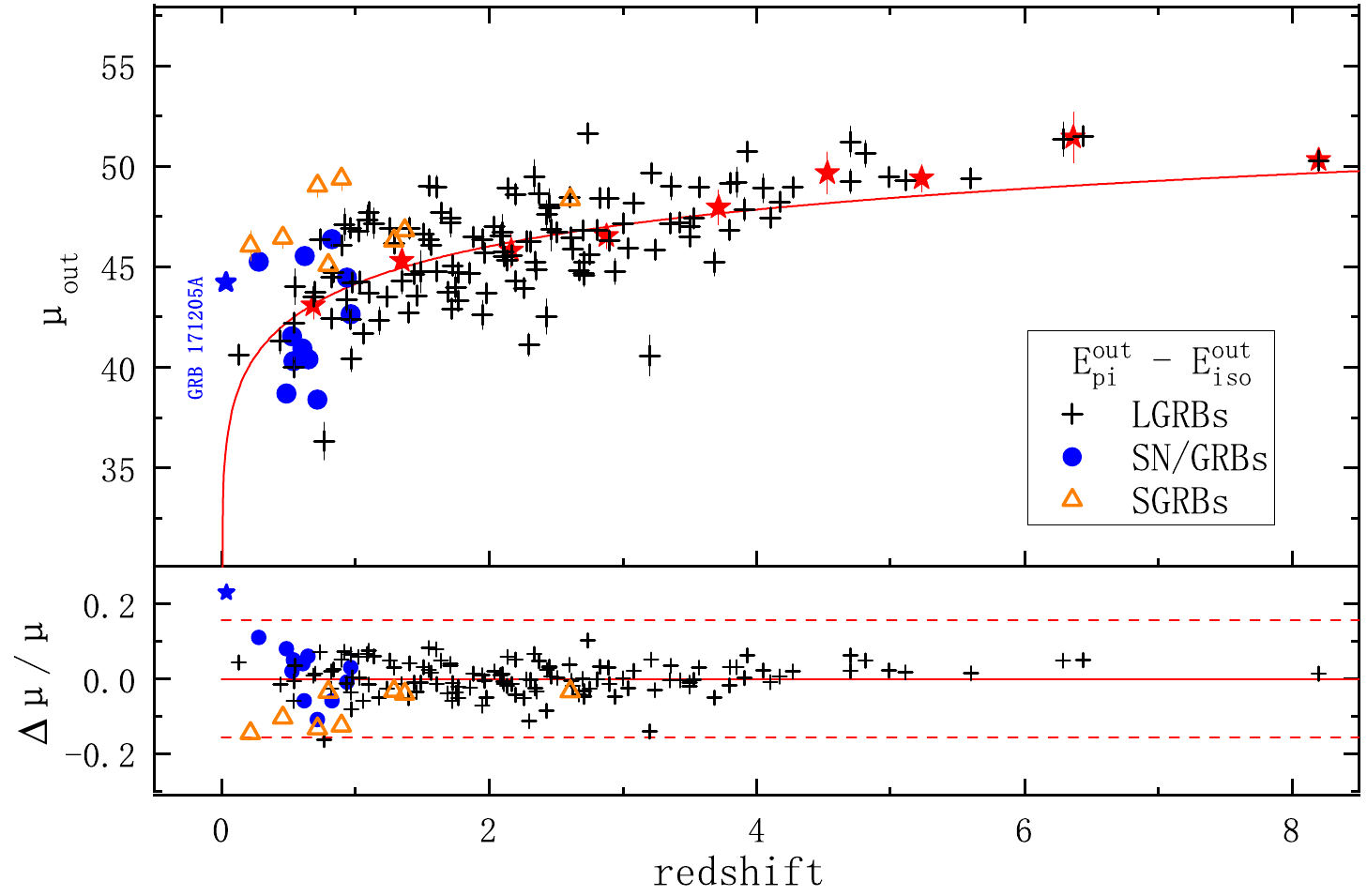}
					\includegraphics[width=0.49\textwidth]{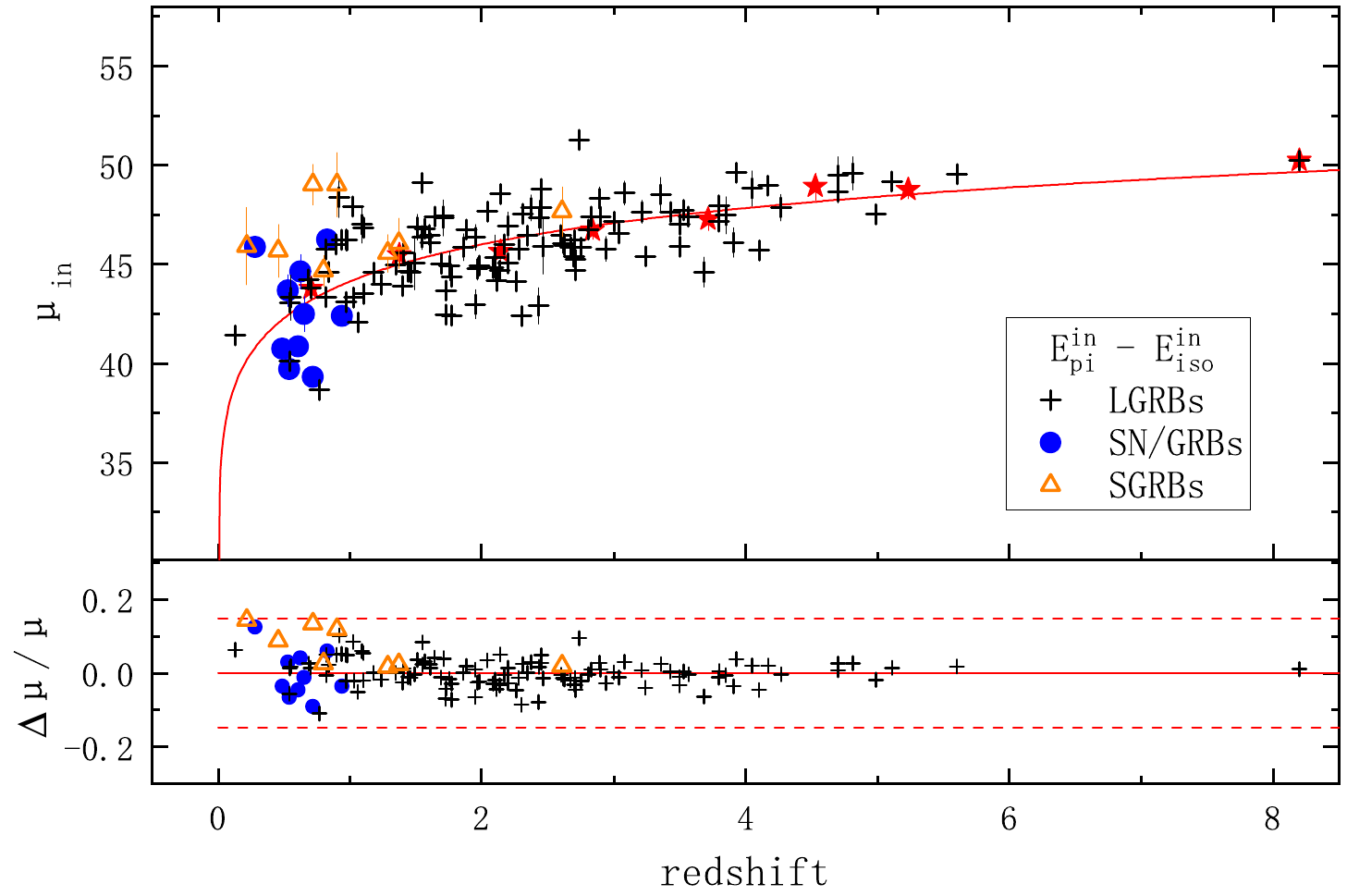}
						\caption{GRB Hubble diagrams in Eq. (\ref{equation:23}) together with relative residuals $\Delta\mu/\mu$ built on the \textbf{$E_{\rm pi}-E_{\rm{iso}}$} relations of out-axis (left panel) and in-axis GRBs (right panel). The black crosses, orange triangles and blue circles respectively stand for normal LGRBs, SGRBs and SN/GRBs. The red stars represent the averaged values of distance module and redshift in each bin. The red solid curves describe the standard cosmology models.
						}
						\label{fig:HD1}
					\end{figure*}
					where $q_{\rm A} (\equiv logC_{\rm A})$ and $\eta_{\rm A}$ can be obtained by
					fitting the $E_{\rm pi}-E_{\rm{iso}}$ relation
					$logE_{\rm pi}=q_{\rm A}+\eta_{\rm A}logE_{\rm iso,52}$ for both out-axis and
					in-axis cases, and the bolmetric fluence $S_{\rm bol}$ is in units
					of $\textrm{erg}\ \textrm{cm}^{-2}$;
					\begin{equation}
						\textrm{(II)} \ \ \ \ \ \ \mu_{GRB}=25+\frac{5}{2}\left[\frac{log E_{\rm pi}-q_{\rm Y}}{\eta_{\rm Y}}-log\left(4\pi P_{\rm bol}\right)+51\right] ,\label{equation:24}
					\end{equation}		
					\begin{figure*}
						\centering
						\footnotesize
						\includegraphics[width=0.49\textwidth]{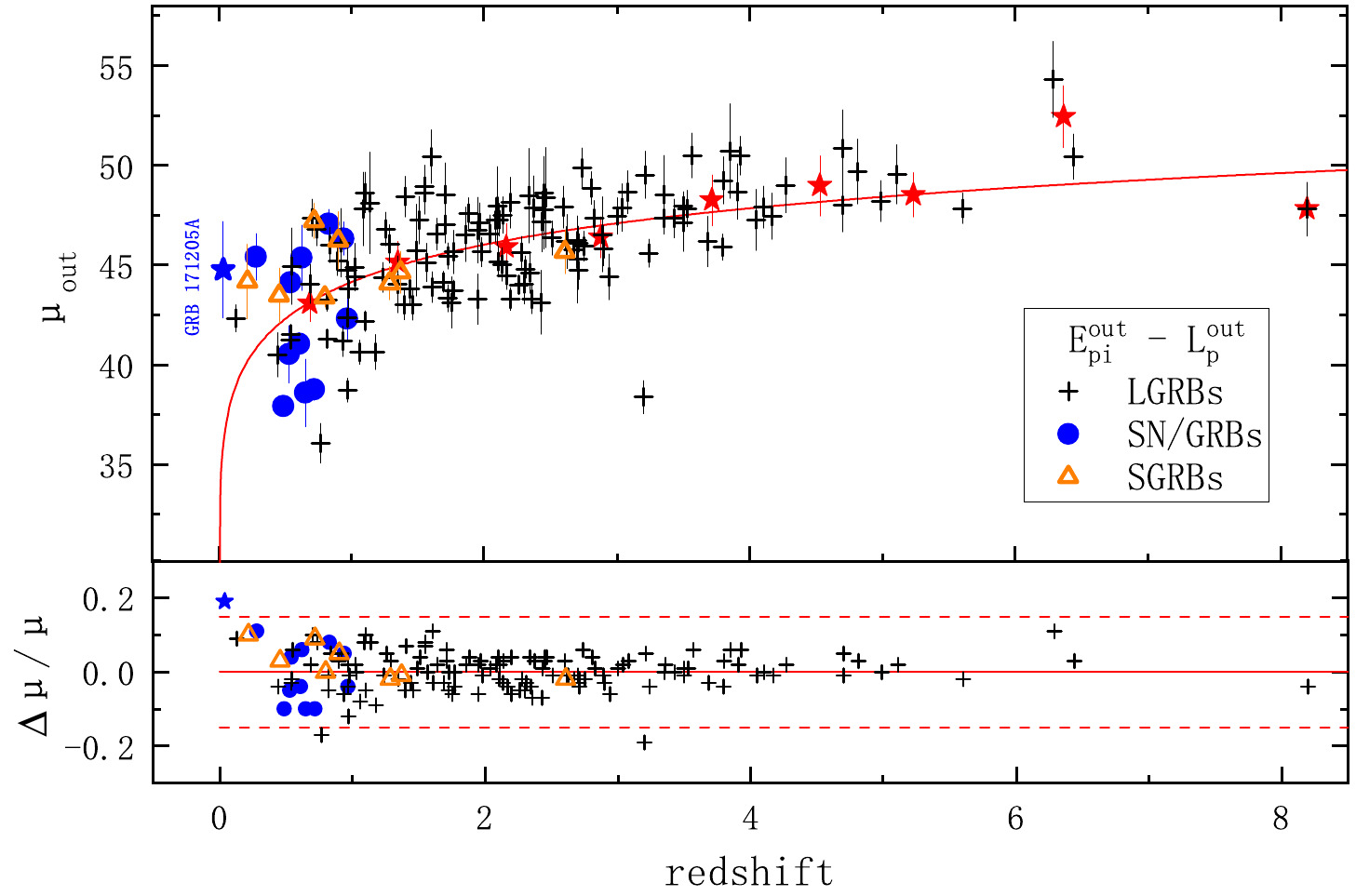}
						\includegraphics[width=0.49\textwidth]{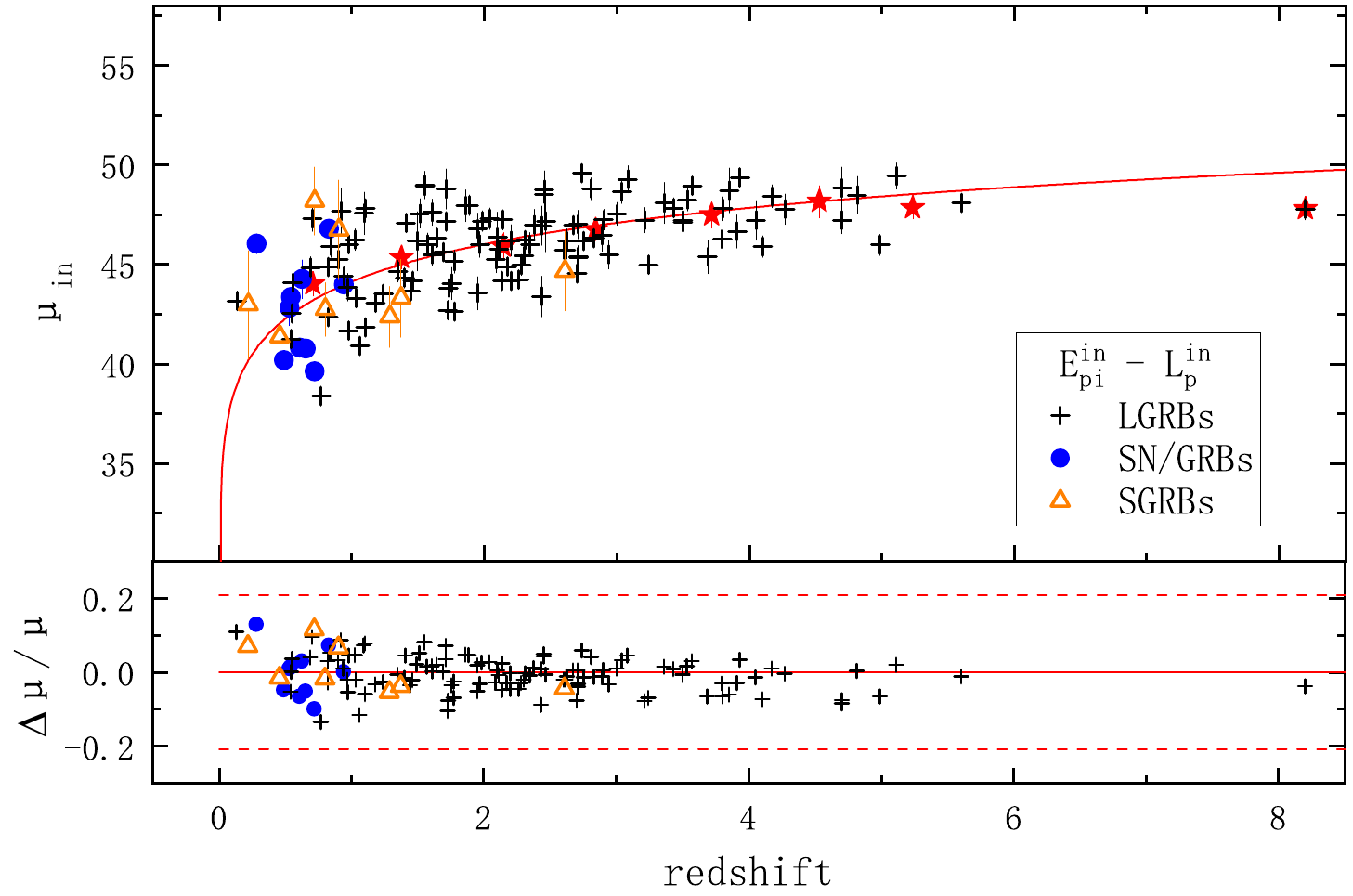}
							\caption{GRB Hubble diagrams in Eq. (\ref{equation:24}) together with relative residuals $\Delta\mu/\mu$ built on the $E_{\rm pi}-L_{\rm{p}}$ relations of out-axis (left panel) and in-axis GRBs (right panel). All symbols are the same as in Figure \ref{fig:HD1}. }
							\label{fig:HD2}
						\end{figure*}	
						where $q_{\rm Y}(\equiv logC_{\rm Y})$ and $\eta_{\rm Y}$ can be similarly
						obtained by fitting the $E_{\rm pi}-L_{\rm{p}}$ relation
						$logE_{\rm pi}=q_{\rm Y}+\eta_{\rm Y}logL_{\rm p,51}$, and the bolometric
						energy flux $P_{\rm bol}$  is in units of $\textrm{erg}\
						\textrm{cm}^{-2}\ \textrm{s}^{-1}$;
						\begin{align}
						 	\notag
							\textrm{(III)} \ \ \ \ \ \ \mu_{GRB}=25+\frac{5}{2}\left[{\frac{log E_{\rm pi}-q_{G}}{\eta_{\rm G}}} \right.\\ \left. -log\left(\frac{4\pi S_{\rm bol}(1-cos\theta_j)}{1+z}\right)+50\right] ,\label{equation:25}
	 					\end{align}
						\begin{figure*}
							\centering
							\footnotesize
							\includegraphics[width=0.49\textwidth]{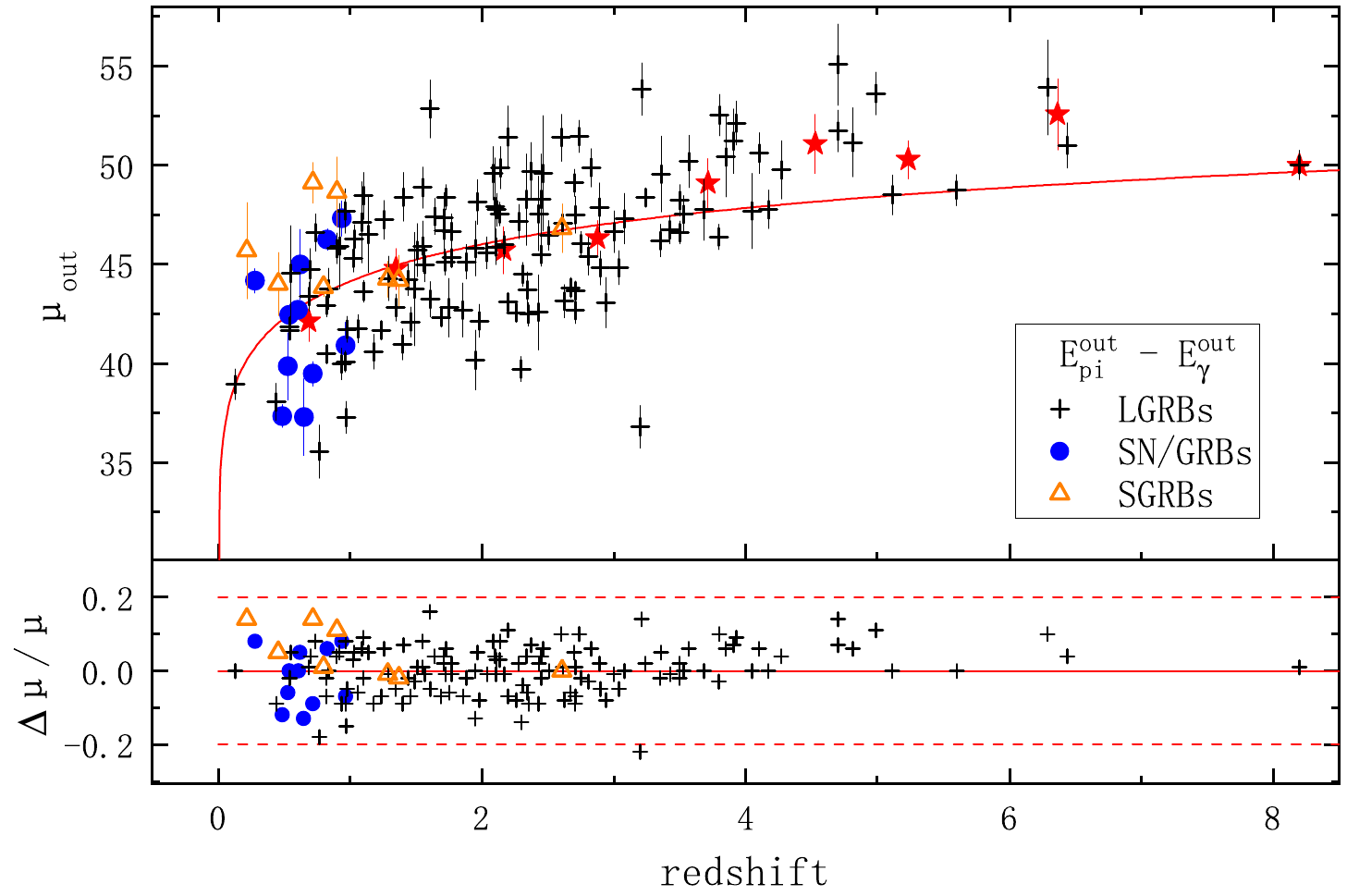}
							\includegraphics[width=0.49\textwidth]{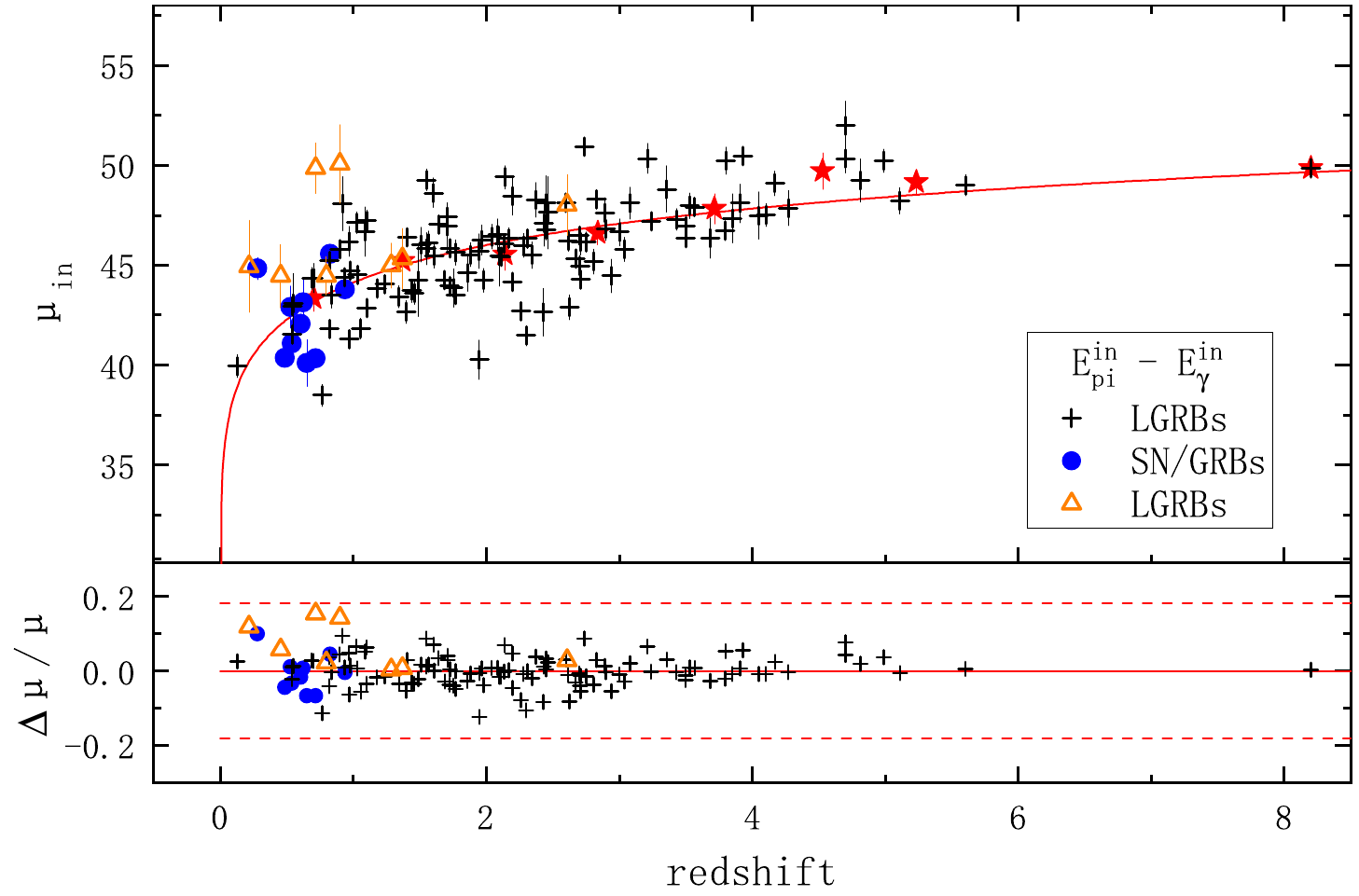}
								\caption{GRB Hubble diagrams in Eq. (\ref{equation:25}) together with relative residuals $\Delta\mu/\mu$ built on the $E_{\rm pi}-E_{\rm{\gamma}}$ relations of out-axis GRBs (left panel) and in-axis GRBs (right panel). All symbols are the same as in Figure \ref{fig:HD1}.}
								\label{fig:HD3}
							\end{figure*}
							in which $q_{G} (\equiv logC_{\rm G})$ and $\eta_{G}$ are derived by fitting the \textbf{$E_{\rm pi}-E_{\rm{\gamma}}$} relation of $logE_{\rm pi}=q_{\rm G}+\eta_{\rm G}\ logE_{\rm \gamma,50}$, with $f_b=1-cos\theta_{\rm j}$ being the beaming	factor. The GRB Hubble diagrams derived from the above	spectrum-energy relations are displayed and compared with theoretical models in Figures \ref{fig:HD1}-\ref{fig:HD3}. In	these plots, the differences between in-axis and out-axis	situations are paid special attention to. Note that the	theoretical distance module of a GRB is determined by $\mu=25+5\		log[D_{l}(z)/(1\textrm{Mpc})]$. Moreover, we divide the data				points into 10 bins and calculate the average distance module and
							redshift in each bin. It is found that the observational GRB
							Hubble diagrams built from both in-axis and out-axis
							spectrum-energy relations of LGRBs match the standard module
							phenomenally. In comparison, the Hubble diagram built on the $E_{\rm pi}-E_{\rm iso}$ relation is better than the other two. At the same time, it should be noted that the three
							in-axis Hubble diagrams are slightly tighter than the corresponding
							out-axis ones. In addition, we emphasize that both SN/GRBs and SGRBs also match the Hubble diagrams with larger scatters although they reside at lower redshifts.
							
							In short, we conclude that while both the in-axis
							and out-axis empirical spectrum-energy relations potentially can
							be used to probe high-redshift universe, the in-axis
							spectrum-energy relations seem to be more precise and are expected
							to present more credible constraints on the cosmological parameters. Our results demonstrate that the larger scatter of previous energy relations could be mainly attributed to the out-axis effect \citep[][Xu23]{2018PASP..130e4202Z} and GRBs have the good perspectives as a standard candle once the out-axis effect is rightly eliminated. It is worth noting that the aim of this study is to simply evaluate the possibility of these in/out-axis GRBs as a probe of cosmological models. However, the circularity problem \citep[e.g.][]{2008MNRAS.391L...1K} is independent of the adoption of the jet model, which may prevent the effective application of these empirical energy relations in cosmology. To overcome the problem, people usually adopt the Markov chain Monte Carlo (MCMC) technique \citep[e.g.][]{2008ApJ...680...92L,2019MNRAS.486L..46A} as an ideal solution to constrain the cosmological parameters with these empirical energy relations combined with other objects jointly \citep{2021Galax...9...77L}. Alternatively, the Bayesian approach should be another real update of the work in the future.


\section{Summary}
							
In this work, we have systematically studied the statistical properties of out-axis GRBs and converted the observed parameters to the in-axis ones. Assuming these out-axis GRBs were produced from structrured power law jets, we have used the out/in-aixs parameters to investigate the three empirical energy relations of peak energy versus isotropic energy, peak luminosity and jet-corrected energy, individually. In addition, we have adopted these newly-built energy relations to build the Hubble diagrams of these out/in-axis GRBs. 	
							
Our main findings are summarized as follows. First, we find that the in-axis average energies are about one order of magnitude larger than the corresponding out-axis values for both short and long GRBs except the Supernova-associated bursts. Second, we show that the $E_{\rm pi}-E_{\rm{iso}}$, $E_{\rm pi}-L_{\rm{P}}$ and $E_{\rm pi}-E_{\rm{\gamma}}$ relations do exist for both out-axis and in-axis GRBs in the framework of the structured jet model. The three energy correlations are respectively $E_{\rm pi}\propto E_{\rm iso}^{0.40}$, $E_{\rm pi}\propto L_{\rm p}^{0.36}$ and $E_{\rm pi}\propto E_{\rm \gamma}^{0.28}$ for the out-axis GRBs, while the energy relations of the in-axis GRBs are $E_{\rm pi}\propto E_{\rm iso}^{0.62}$, $E_{\rm pi}\propto L_{\rm p}^{0.57}$ and $E_{\rm pi}\propto E_{\rm \gamma}^{0.47}$ correspondingly. It can be found that the in-axis energy relations become steeper than those corresponding out-axis relations universally. Third, our power-law indices in energy relations of in/out-axis LGRBs are consistent with the values of on/off-axis LGRBs derived by X23 on basis of Synchrotron radiation mechanism, which confirms that the Synchrotron radiation mechanism should dominate the in/out-axis bursts. The slight fluctuations could be caused by the discrepancy of jet geometry. Fourth, we utilize these newly-found energy relation of out/in-axis GRBs to build their Hubble diagrams. It is found that the Hubble diagrams of in-axis GRBs are tighter than those of out-axis bursts and can be applied as good cosmological tools. Our results demonstrate that the larger scatter of previous energy relations could be mainly attributed to the out-axis effect and GRBs have the good perspectives as a standard candle once the out-axis effect is correctly eliminated no matter what kinds of jet dynamics are considered.

\section*{Acknowledgements}

We are very thankful to the referee for his/her constructive suggestions and
comments. This work was supported in part by National Natural Science Foundation of China
(grant Nos. 12588202, U2031118, 11873030, 12041306, U1938201, 12233002), the Youth Innovations and Talents
Project of Shandong Provincial Colleges and Universities (grant No. 201909118), the National Key R\&D
Program of China (2021YFA0718500), the National SKA Program of China No. 2020SKA0120300 and the
Natural Science Foundations (ZR2018MA030, XKJJC201901). YFH also acknowledges the support from the Xinjiang Tianchi Program.
\bibliography{sample631}{}
\bibliographystyle{aasjournal}




\end{document}